\documentclass{aa}
\usepackage{psfig}

\def\h2{H{\small II}}
\newcounter{qub}
\begin{document}
%  \thesaurus{  03          % Extragalactic Astronomy
%              (11.01.1;    % Galaxies: abundances
%               11.04.2;    % Galaxies: dwarf
%               11.05.2;    % Galaxies: evolution
%               11.03.2;    % Galaxies: compact
%               11.19.3;    % Galaxies: starburst
%               11.19.5; )} % Galaxies: stellar content

\title{Deep VLT spectroscopy of the
blue compact dwarf galaxies Tol~1214--277 and 
Tol~65\thanks{Based on observations
obtained at the European Southern Observatory, Paranal, Chile 
(ESO Programs 63.P-0003 and 65.N-0642)}}

%
%\subtitle{Constraint on age.}
%

\author{Y. I. Izotov \inst{1}
\and P. Papaderos \inst{2}
\and N. G. Guseva \inst{1}
\and K. J. Fricke \inst{2}
\and T. X. Thuan\inst{3}}
\offprints{izotov@mao.kiev.ua}
\institute{      Main Astronomical Observatory,
                 Ukrainian National Academy of Sciences,
                 27 Zabolotnoho str., Kyiv 03680,  Ukraine
\and
                 Universit\"ats--Sternwarte, Geismarlandstra\ss e 11,
                 D--37083 G\"ottingen, Germany
\and
                 Astronomy Department, University of Virginia, 
                 Charlottesville, VA 22903, USA
}

\date{Received \hskip 2cm; Accepted}

\abstract{We present VLT spectroscopic observations with different
spectral resolutions and different slit orientations of the two
metal-deficient blue compact dwarf (BCD) galaxies Tol 1214--277 and Tol 65.
The oxygen abundances in the brightest H {\sc ii} regions of Tol 1214--277 and 
Tol 65 are found to be 12 + log O/H = 7.55 $\pm$ 0.01 and 7.54 $\pm$ 0.01, 
or $Z_\odot$/24\thanks{12+log(O/H)$_{\odot}$ = 8.92
(Anders \& Grevesse \cite{Anders89}).}. The nitrogen-to-oxygen
abundance ratios in the two galaxies are log N/O = --1.64 $\pm$ 0.03 and 
--1.60 $\pm$ 0.02 and lie in the narrow 
range found for other most metal-deficient BCDs. 
The helium mass fraction derived in several H {\sc ii} regions in both 
galaxies is consistent with a high primordial helium
mass fraction, $Y_{\rm p}$ $\sim$ 0.244.
We confirm the detection of the high-ionization forbidden emission line 
[Fe {\sc v}] $\lambda$4227 in the spectrum of Tol 1214--277.
% which suggests the hard radiation field likely 
% produced by the combined effect of massive O stars and radiative shocks. 
Additionally, weak [Ne {\sc iv}] $\lambda$4725, [Fe {\sc vi}] $\lambda$5146, 
$\lambda$5177, and
[Fe {\sc vii}] $\lambda$5721, $\lambda$6087 emission lines are detected in 
the high-resolution spectrum of Tol 1214--277. The detection of these lines
implies the presence of hard radiation with photon energy in the range 
$\sim$ 4 -- 8 Ryd. 
%Eight emission-line galaxies have been detected 
Emission lines are detected in the spectra of eight galaxies 
in the fields of Tol 1214--277 and Tol 65. Seven of these galaxies are 
background objects,
while one galaxy has a redshift close to that of Tol 1214--277. 
Situated at a projected distance of $\sim$ 14.5 kpc from Tol 1214--277, this 
galaxy is probably a companion of the BCD.\thanks{Tables 2, 3, 4 and 8 are also 
available in electronic form at the CDS via anonymous ftp to 
cdsarc.u-strasbg.fr (130.79.128.5) or via 
http://cdsweb.u-strasbg.fr/cgi-bin/qcat?J/A+A/}
}

\titlerunning {Deep VLT spectroscopy of Tol 1214--277 and Tol 65}

\maketitle

\section{Introduction}

%*********************************************************
%*********************************************************
%  Fig.1 - the slit orientations 
%*********************************************************
%*********************************************************

\begin{figure*}%[hbtp]
    \hspace*{0.0cm}\psfig{figure=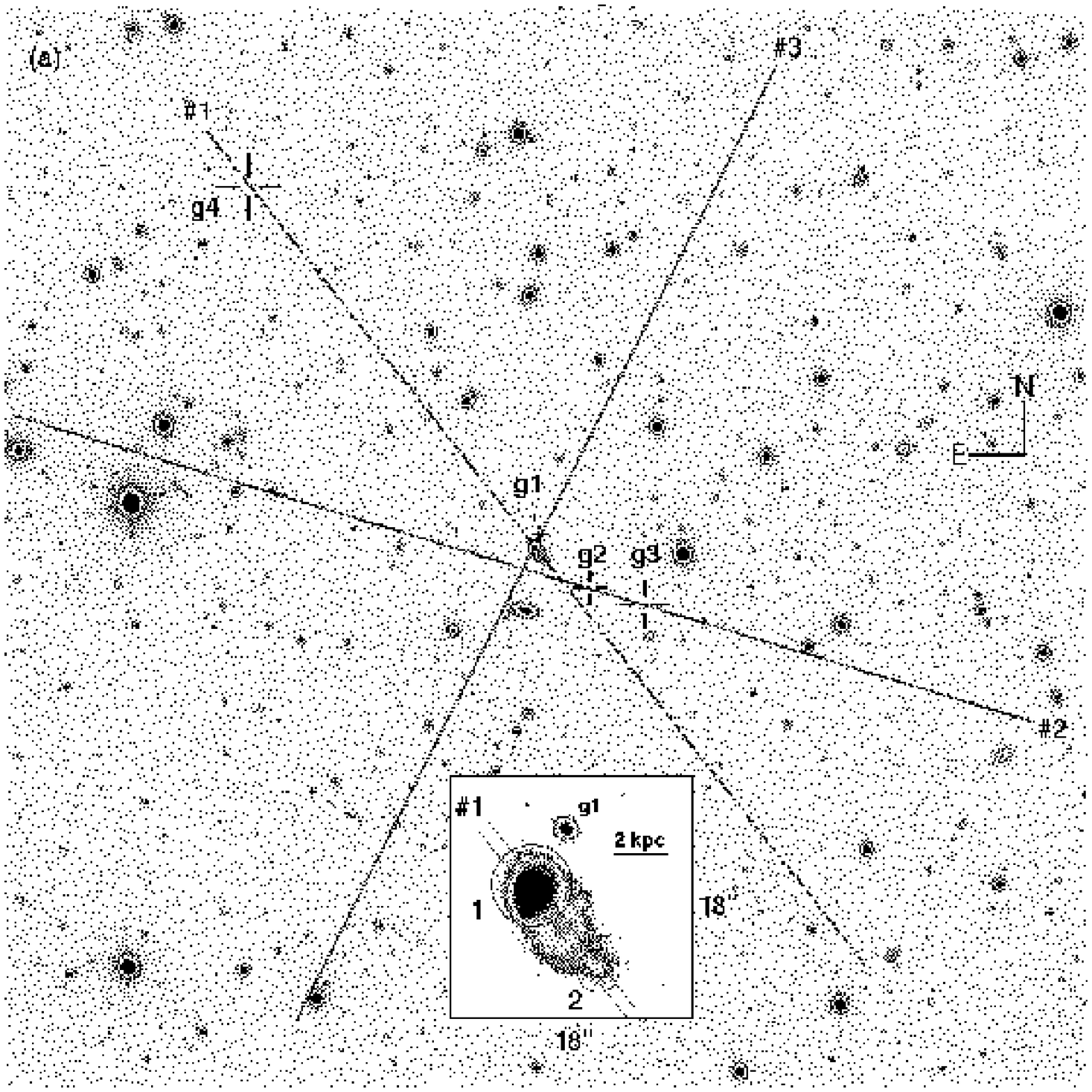,angle=0,width=8.9cm,clip=}
    \hspace*{0.05cm}\psfig{figure=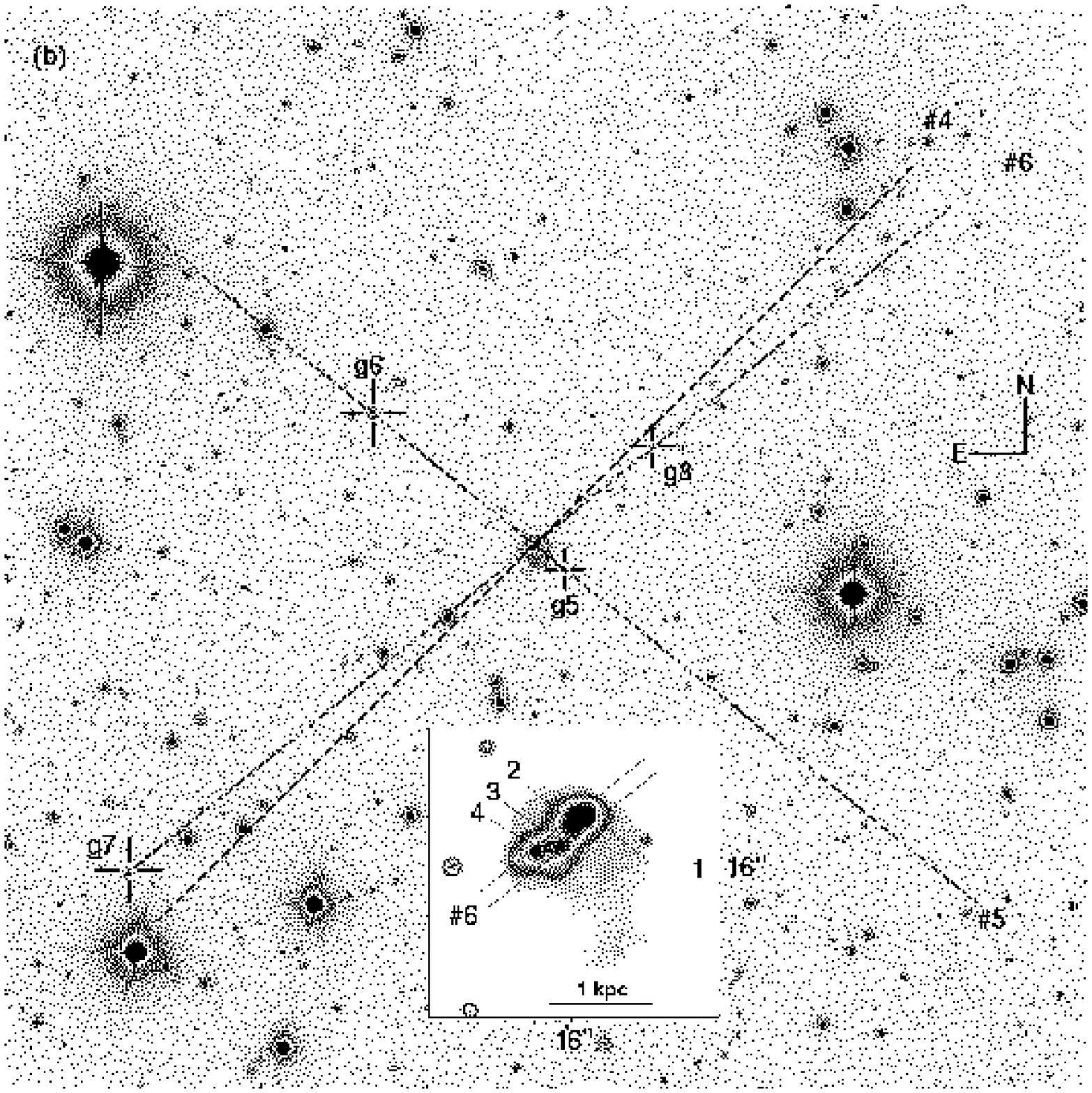,angle=0,width=8.9cm,clip=}
    \caption{Tol 1214--277 (left panel) and Tol 65 (right panel)
centered on the VLT FORS\,1 6\farcm8$\times$6\farcm8 field of view. 
The positions of the slits used for 
spectroscopic observations are shown by straight lines and are 
labeled \#1 to \#3 and \#4 to \#6, respectively.
The insets display enlarged versions of the BCDs with the slits 
\#1 and \#6 overlayed on Tol 1214--277 and Tol 65, respectively.
The H {\sc ii} regions 1 and 2 in Tol 1214--277
(left panel) and 1 to 4 in Tol 65 (right panel) are indicated.
%North is up and east is to the left.
}
    \label{fig1}
\end{figure*}

%*******************************
% Table 1
%*******************************
\begin{table*}%[h]%[tbh]
\caption{Journal of observations.}
\label{tab1}
\begin{tabular}{lcclccrccc} \hline \hline
Field       & Date     &Slit \#&Spectral elements&Spectral range   &Slit    & P.A.         &Exposure  &Airmass&Seeing    \\ \hline
Tol 1214--277&18.05.1999&  1    &G300V + GG 375   &3600--7400\AA    &1\arcsec& 39$^\circ$ & 2$\times$660 s    & 1.7   &0\farcs69 \\
             &25.06.2000&  1    &G300V + GG 375   &3600--7400\AA    &1\arcsec& 39$^\circ$ & 2$\times$900 s    & 1.5   &1\farcs12 \\
             &28.06.2000&  2    &G300V + GG 375   &3600--7400\AA    &1\arcsec& 73$^\circ$ & 2$\times$900 s    & 1.5   &1\farcs04 \\
             &25.06.2000&  3    &G600B            &3600--5400\AA    &1\arcsec&154$^\circ$ & 3$\times$900 s    & 1.1   &1\farcs33 \\
             &25.06.2000&  3    &G600R + GG 435   &5300--7400\AA    &1\arcsec&154$^\circ$ & 2$\times$900 s    & 1.2   &1\farcs61 \\
Tol 65       &15.05.1999&  4    &G300V + GG 375   &3600--7400\AA    &1\arcsec&135$^\circ$ & 2$\times$420 s    & 1.4   &0\farcs34 \\
             &30.06.2000&  5    &G300V + GG 375   &3600--7400\AA    &1\arcsec& 51$^\circ$ & 2$\times$900 s    & 1.6   &0\farcs80 \\
             &30.06.2000&  6    &G600B            &3600--5400\AA    &1\arcsec&129$^\circ$ & 3$\times$900 s    & 1.2   &0\farcs96 \\
             &30.06.2000&  6    &G600R + GG 435   &5300--7400\AA    &1\arcsec&129$^\circ$ & 2$\times$900 s    & 1.3   &1\farcs03 \\ \hline
\end{tabular}
\end{table*}

%*********************************************************
%*********************************************************
%  Fig.2 - high resolution spectrum of Tol 1214-277
%*********************************************************
%*********************************************************

\begin{figure*}%[hbtp]
    \hspace*{3.cm}\psfig{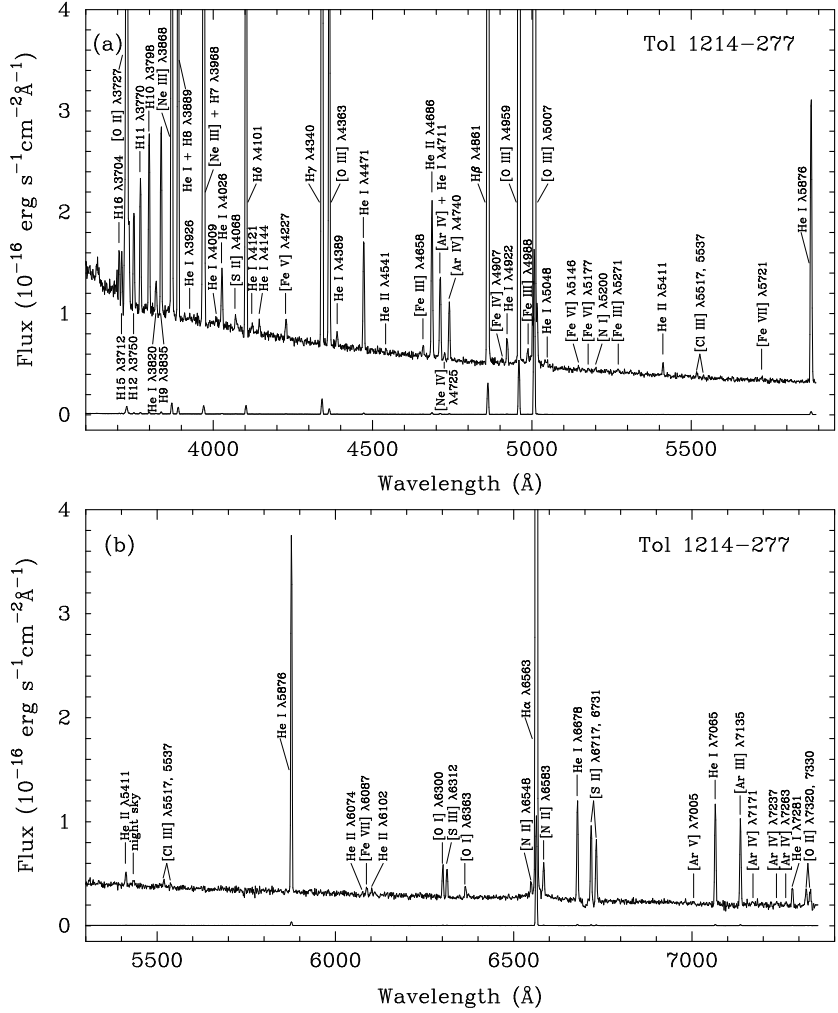}
    \caption{Blue (a) and red (b) parts of the 
redshift-corrected high-resolution spectrum of 
the brightest H {\sc ii} region (region 1) in Tol 1214--277 obtained in 2000 
with slit position \#3. The emission lines are labeled.
The lower spectra in (a) and (b) are the observed spectra 
downscaled by a factor of 100.}
    \label{fig2}
\end{figure*}

%*********************************************************
%*********************************************************
%  Fig.3 - high resolution spectrum of Tol 65 (regions 1+2)
%*********************************************************
%*********************************************************

\begin{figure*}%[hbtp]
    \hspace*{3.cm}\psfig{figure=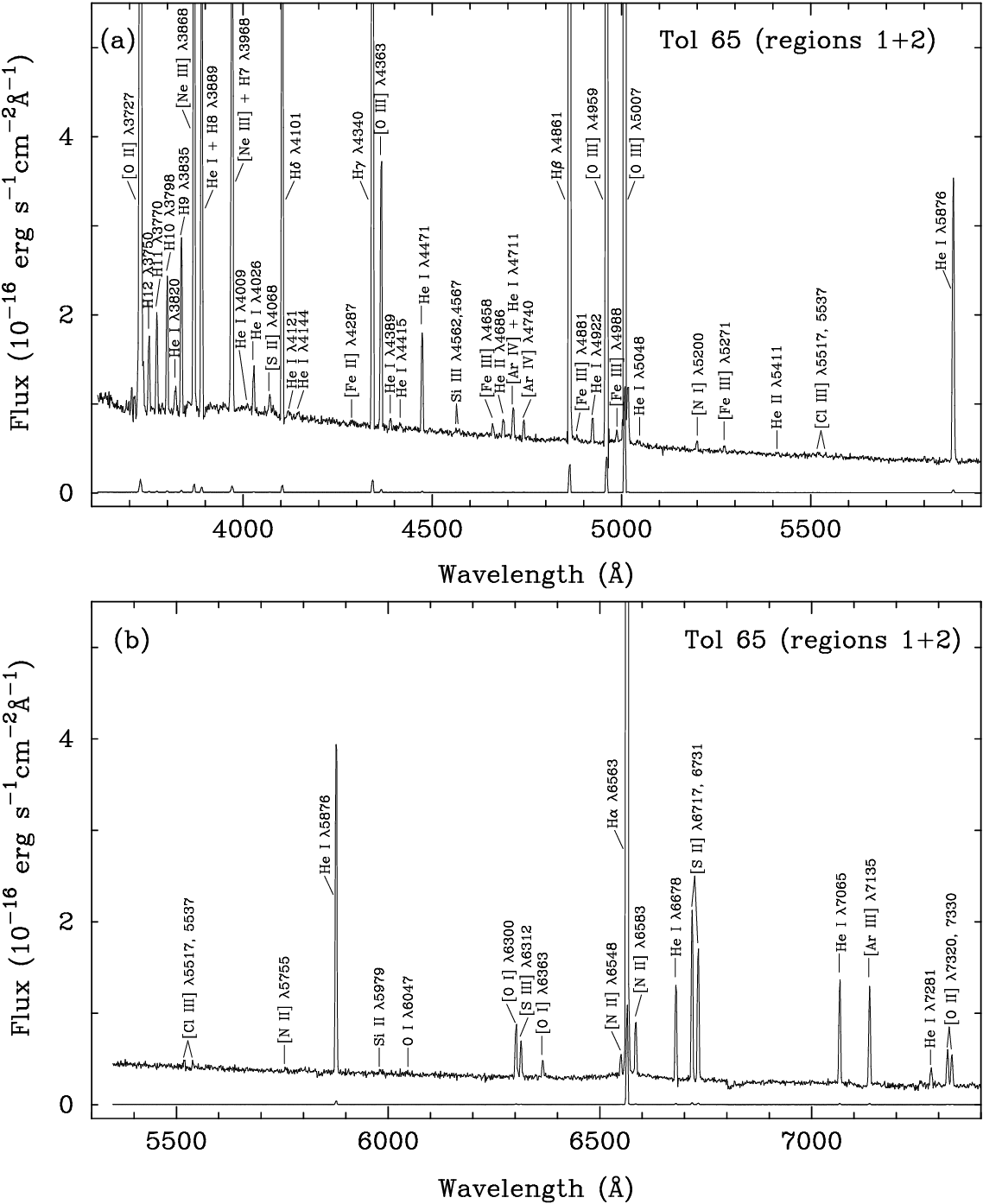,angle=0,width=11.cm,clip=}
    \caption{Blue (a) and red (b) parts of the redshift-corrected 
high-resolution spectrum of regions 1 + 2 in Tol 65 obtained in 2000 with slit 
position \#6. The emission lines are labeled.
The lower spectra in (a) and (b) are the observed spectra 
downscaled by a factor of 100.}
    \label{fig3}
\end{figure*}

%*********************************************************
%*********************************************************
%  Fig.4 - high resolution spectrum of Tol 65 (regions 3+4)
%*********************************************************
%*********************************************************

\begin{figure*}%[hbtp]
    \hspace*{3.cm}\psfig{figure=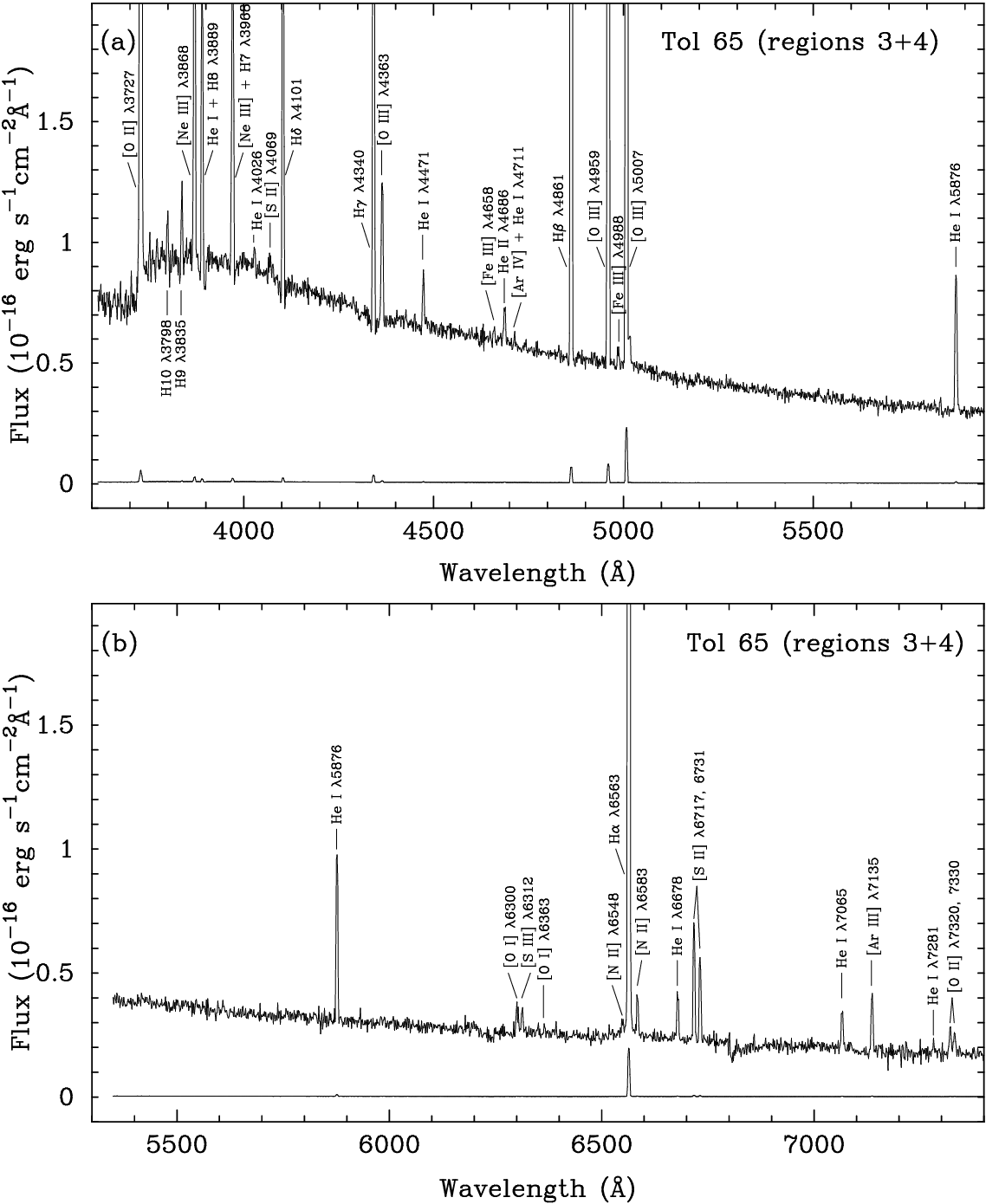,angle=0,width=11.cm,clip=}
    \caption{Blue (a) and red (b) parts of the redshift-corrected 
high-resolution spectrum of 
regions 3 + 4 in Tol 65 obtained in 2000 with slit 
position \#6. The emission lines are labeled.
The lower spectra in (a) and (b) are the observed spectra 
downscaled by a factor of 100.}
    \label{fig4}
\end{figure*}

%*********************************************************
%*********************************************************
%  Fig.5 - low resolution spectra of Tol 1214-277
%*********************************************************
%*********************************************************

\begin{figure*}%[t]
    \hspace*{3.cm}\psfig{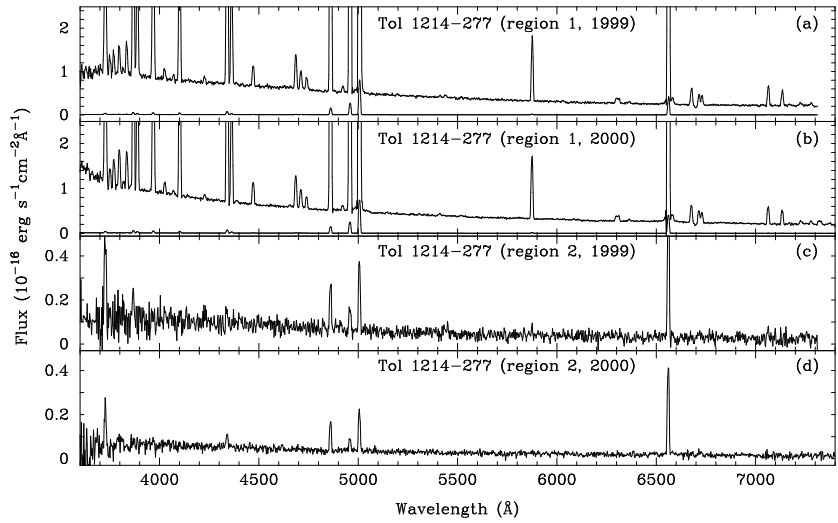}
    \caption{Redshift-corrected low-resolution spectra of regions 1 and 2 in 
Tol 1214--277 obtained in 1999 and 2000 with slit position \#1.
The lower spectra in (a) and (b) are the observed spectra downscaled by a 
factor of 100.}
    \label{fig5}
\end{figure*}

%*********************************************************
%*********************************************************
%  Fig.6 - low resolution spectra of Tol 65
%*********************************************************
%*********************************************************

\begin{figure*}%[hbtp]
    \hspace*{3.cm}\psfig{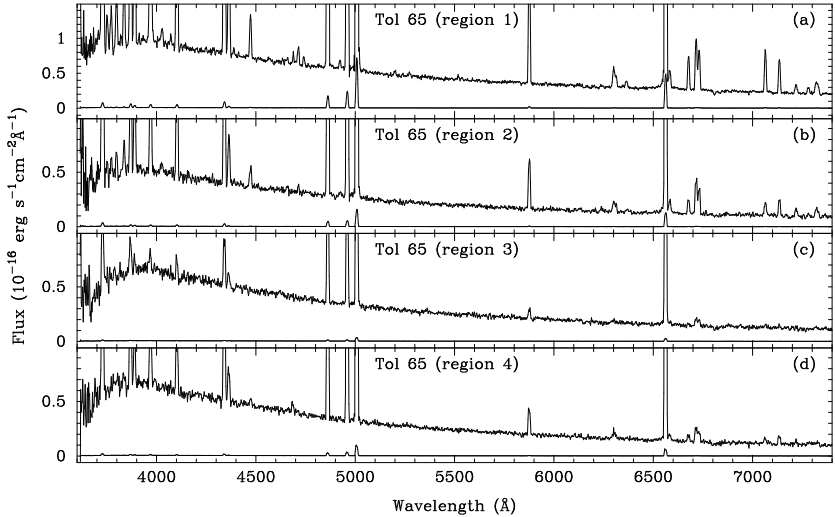}
    \caption{Redshift-corrected low-resolution spectra of regions 1 -- 4 in 
Tol 65 obtained in 1999 with slit position \#4.
The lower spectra are the observed spectra downscaled by a 
factor of 100.}
    \label{fig6}
\end{figure*}

Detailed spectroscopic studies of extremely metal-poor local blue compact 
dwarf (BCD) galaxies based on observations with large telescopes are useful 
to derive accurately their element abundances and shed 
light on the properties and origin of these galaxies. 
BCDs may be considered local counterparts to high-redshift nearly primordial
star-forming galaxies because of their low metallicity and active star 
formation. Spectroscopic studies
of other galaxies in the field of BCDs are aimed at searching for
companion galaxies. Such a companion galaxy has been found, 
e.g. for the second most metal-deficient BCD known, SBS 0335--052, 
at a projected distance of $\sim$ 22 kpc
(Pustilnik et al. \cite{PU97,PU01}; Lipovetsky et al. \cite{L99}). The 
gravitational interaction between the galaxies in such pairs has been 
suggested as a triggering
mechanism of star formation in BCDs (Pustilnik et al. \cite{PU01};
Noeske et al. \cite{N01}).

We focus here on two BCDs, Tol 1214--277 $\equiv$ Tol 21 
and Tol 65 $\equiv$ Tol 1223--359 
$\equiv$ ESO 380 -- G027. Using deep VLT imaging of these
BCDs, Fricke et al. (\cite{F01}) and Papaderos et al. (\cite{P99}) have 
studied their morphological properties. In Tol 1214--277, a very bright and 
compact star-forming region 
(region 1 in the inset of Fig. \ref{fig1}, left panel) 
at the north-eastern edge of the galaxy is embedded in a 
relatively blue extended low-surface-brightness (LSB) component with some 
other fainter star-forming regions (Fricke et al. \cite{F01}). 
As for Tol 65, Papaderos et al. (\cite{P99}) have found 
several regions of star formation in the northern part (see inset 
in  Fig. \ref{fig1}, right panel), also embedded in a relatively blue LSB 
component. The distances to Tol 1214--277 and Tol 65 are respectively 
103.9 Mpc and 36.0 Mpc, using the observed 
radial velocities of 7795 km s$^{-1}$ and 2698 km s$^{-1}$
and assuming a Hubble constant $H_0$ = 75 km s$^{-1}$ Mpc$^{-1}$
(Thuan \& Izotov \cite{TI97}).
The presence of massive stars in these BCDs is indicated by strong 
nebular lines (Izotov et al. \cite{I01a}) and, in the case of Tol 1214--277, 
by a UV stellar N {\sc v} $\lambda$1240 line with a P Cygni profile 
(Thuan \& Izotov \cite{TI97}).

The very low metallicity of Tol 1214--277 and Tol 65
has been established by earlier spectroscopic work 
(Kunth \& Sargent \cite{K83}; Campbell, Terlevich \& Melnick \cite{C86}; 
Pagel et al. \cite{P92}; Masegosa, Moles \& Campos-Aguilar \cite{M94}). 
Recently, Fricke et al. (\cite{F01}) and Izotov et al. (\cite{I01a}) 
using VLT and Keck spectroscopic observations have derived 
the oxygen abundance 12 + log O/H = 7.52 and 7.54 in Tol 1214--277,
and Izotov et al. (\cite{I01a}) have found 12 + log O/H = 7.54 in Tol 65.
Hence these two galaxies are the most metal-deficient BCDs known 
in the Southern hemisphere, after SBS 0335--052 with 12 + log O/H = 7.30 
(Izotov et al. \cite{I97b}). 
We present here new VLT spectroscopic observations of these two BCDs to 
pursue two problems. The first one concerns the N/O abundance ratio. 
Pagel et al. (\cite{P92}) have measured the nitrogen-to-oxygen abundance 
ratio in Tol 65 to be log N/O = --1.81 $\pm$ 0.15, significantly  
lower than the mean value log N/O $\approx$ --1.60 $\pm$ 0.02 obtained by 
Thuan et al. (\cite{TIL95}) and Izotov \& Thuan (\cite{IT99})
for low-metallicity BCDs with 12 + log O/H $\la$ 7.6. On the other hand, 
the N/O abundance ratio in Tol 1214--277 is similar to that of other BCDs.
 Recently,
Izotov et al. (\cite{I01a}) using low-resolution Keck spectra of Tol 1214--277 and 
Tol 65 have found log N/O = --1.64 for both galaxies, more in line with other 
low-metallicity BCDs.
To understand the origin of nitrogen in a low-metallicity 
environment (e.g., Meynet \& Maeder \cite{MM02}), it is important to check 
these results with  
new high-quality spectroscopic data with better spectral resolution. 

The second problem concerns the nature of the hard radiation field 
in low-metallicity BCDs. Tol 1214--277 shows the strongest 
He {\sc ii} $\lambda$4686 emission among all known 
star-forming emission-line galaxies. This suggests that the radiation field in
the BCD is very hard, with a significant amount of the ionizing photons at
$\lambda$ $<$ 228\AA, equivalent to a photon energy $>$ 4 Ryd. 
Fricke et al. (\cite{F01}) have discovered the high-ionization 
[Fe {\sc v}] $\lambda$4227 emission line in the spectra of Tol 1214--277
and SBS 0335--052, further supporting the presence of
hard radiation in the star-forming regions of low-metallicity BCDs. 
They have proposed that the main source of
radiation at $\lambda$ $<$ 228\AA\ are fast shocks. Later
Izotov, Chaffee \& Schaerer (\cite{I01b}) confirmed the presence of 
the [Fe {\sc v}] $\lambda$4227 emission line in the spectrum of SBS 0335--052 
and found some other high-ionization 
[Fe {\sc v}] -- [Fe {\sc vii}] emission lines. 
The source of highly ionized iron
is not yet well established and new high signal-to-noise spectroscopic 
observations are necessary to understand the main sources of high ionization.

In the following, we investigate the 
spectral properties of Tol 1214--277 and Tol 65 with new deep VLT data.
In Sect. \ref{observ} we describe the data, and in Sect. \ref{chem} we discuss
the chemical abundances.
High-ionization emission lines and the origin of hard radiation in 
Tol 1214--277 are discussed in Sect. \ref{highion}. 
The properties of the emission-line galaxies in the fields of
Tol 1214--277 and Tol 65 are discussed in Sect. \ref{field}. We summarize 
our results in Sect. \ref{sum}.

% =======================================
\section{Observations and data reduction \label{observ}}
% =======================================

Spectroscopic data for Tol 1214--277 and Tol 65 with different position 
angles (P.A.) were taken with VLT/FORS 1 in service mode 
during two observing periods, on 15 -- 18 May, 1999 and
25 -- 30 June, 2000 (Table \ref{tab1}). The positions of the slits are
shown in Fig. \ref{fig1}. Low-resolution spectra were
obtained with the G300V grism and the GG 375 second-order blocking filter.
This yields a spatial resolution along the slit of 0\farcs2 pixel$^{-1}$,
a scale perpendicular to the slit of $\sim$ 2.6 \AA\ pixel$^{-1}$, a spectral
coverage of 3600 -- 7400 \AA, and a spectral resolution of $\sim$ 10 \AA\ (FWHM).
High-resolution spectra have been obtained separately for the blue and
red parts with the grisms G600B and G600R respectively, giving 
a spatial resolution along 
the slit of 0\farcs2 pixel$^{-1}$, a scale perpendicular to the slit of 
$\sim$ 1.2 \AA\ pixel$^{-1}$, and a spectral resolution of $\sim$ 5 \AA\ (FWHM).
The spectra in the red parts were obtained 
with a GG 435 second-order blocking filter. 
The spectral coverage was 3600 -- 5400 \AA\ in the blue 
and 5300 -- 7400 \AA\ in the red. A 1\arcsec\ $\times$ 400\arcsec\ slit 
was used for all observations. The airmass during the observations varied from
1.1 to 1.7. The positions of 
the slits were chosen to go through all brightest H {\sc ii} regions in
both BCDs and some suspected emission-line galaxies in the fields of
Tol 1214--277 and Tol 65 (Fig. \ref{fig1}, Table \ref{tab1}).
With the total exposure times shown in Table \ref{tab1}, the spectra have
a signal-to-noise ratio S/N $\ga$ 50 in the continuum of the 
brightest parts of the BCDs. They were broken up into two 
subexposures, to allow for cosmic-ray removal. 
A spectrum of a He-Ne-Ar comparison lamp was obtained 
for wavelength calibration. 

Data reduction was carried out using the 
IRAF\footnote{IRAF: the Image Reduction
and Analysis Facility is distributed by the National Optical Astronomy 
Observatory.} software package. 
This included bias subtraction, cosmic-ray removal,
flat-field correction, wavelength calibration and night-sky emission 
subtraction.
Spectrophotometric standard stars were observed during the same nights
with 2\arcsec\ wide slits.
However, because of hardware problems of the spectrograph those spectra were 
shifted by $\sim$ 300 pixels along the dispersion axis compared to
object spectra, and were therefore unusable. 
The flux calibration was done using 
calibrated spectra of Tol 1214--277 and Tol 65 obtained previously by 
Izotov et al. (\cite{I01a}) with the Keck telescope. For this, we 
removed the emission lines and obtained the continuum distributions for
both the flux-calibrated Keck and uncalibrated VLT spectra.
The transformation curve to flux-calibrate the VLT spectra is then
obtained by dividing one continuum by the other.

\section{Chemical abundances \label{chem}}

We derive the element abundances in the brightest H {\sc ii}
regions of Tol 1214--277 and Tol 65 using all available data obtained during 
the two periods of observations. 
The seeing during the observations varied over a wide 
range, from 0\farcs34 to 1\farcs61 (FWHM). We were able to extract 
one-dimensional spectra of all four 
regions from the low-resolution long-slit spectrum of Tol 65, obtained
on 15 May 1999 (slit \#4) with exceptionally good seeing (0\farcs34 FWHM). 
However, 
the seeing during the high-resolution spectroscopic observations of Tol 65
on 30 June, 2000 (slit \#6) was not as good (1\farcs03 FWHM), so that 
we were able
to extract only one-dimensional spectra for the combined regions (1+2) and 
(3+4). The one-dimensional spectra from the low-resolution long-slit 
spectrum of Tol 65 were extracted within an aperture of 
1\arcsec\ $\times$ 1\arcsec. In all other cases, the spectra were extracted 
within an aperture of 2\arcsec\ $\times$ 1\arcsec.

The emission line fluxes were measured with the IRAF SPLOT 
routine using Gaussian profile fitting. 
The errors of the line fluxes are defined by the total photon counts in
each line and are derived from the non-flux-calibrated spectra.
They have been propagated in the calculations of the elemental abundance 
errors.
Fluxes have been corrected for underlying stellar absorption (for hydrogen 
lines) and interstellar extinction using the observed Balmer decrement,
following the procedure described by Izotov et al. (\cite{ITL94,ITL97}).
This is done by minimizing the deviations of the hydrogen emission
line fluxes corrected for extinction and underlying absorption, from their
theoretical values. All non-blended hydrogen emission lines were used. The 
Whitford (\cite{W58}) reddening law was adopted. 

\subsection{Emission-line fluxes}

\subsubsection{High-resolution spectra}
 
The spectrum of the brightest region 1 of Tol 1214--277 was obtained with slit
\#3 (Fig. \ref{fig1}), oriented roughly perpendicular to the major axis of the 
galaxy. It is shown in Fig. \ref{fig2}. The spectrum is dominated by very 
strong emission lines, reflecting the ongoing star formation activity. 
Because of the high signal-to-noise ratio and the high spectral resolution, 
several weak permitted and forbidden nebular emission lines 
are present in the spectrum. The emission lines [O {\sc i}] $\lambda$6300 and 
[S {\sc iii}] $\lambda$6312, H$\alpha$ $\lambda$6563 and [N {\sc ii}] $\lambda$6583
are well separated, allowing 
for a reliable determination of sulfur and nitrogen abundances.
Remarkable spectral features are the nebular 
high-ionization He {\sc ii} $\lambda$4686,
$\lambda$5411 and [Fe {\sc v}] $\lambda$4227 emission lines
discussed by Fricke et al. (\cite{F01}) and suggesting the presence of 
a very hard radiation with photon energies $\ga$ 4 Ryd in the BCD.
In addition to these findings, we detect the [Ne {\sc iv}] $\lambda$4725 
emission line and weak [Fe {\sc vi}]
$\lambda$5146, $\lambda$5177, [Fe {\sc vii}] $\lambda$5721, $\lambda$6087  
emission lines. 
This makes Tol 1214--277 the second known BCD, after
SBS 0335--052 (Izotov et al. \cite{I01b}), with detected [Fe {\sc vi}] -- 
[Fe {\sc vii}] emission lines. 

The corrected fluxes $I(\lambda)/I({\rm H}\beta)$ and equivalent widths 
$EW$ of emission lines together with the extinction coefficient $C$(H$\beta$), 
the absolute flux $F$(H$\beta$) of the H$\beta$ emission line and 
the equivalent width $EW$(abs) of hydrogen absorption lines 
for region 1 of Tol 1214--277 are listed in Table \ref{tab2}. 

The high resolution spectra of regions 1+2 and 3+4 in Tol 65 were obtained with
slit \#6 (Fig. \ref{fig1}) oriented along the chain of star-forming regions 
in the northern part of the galaxy and are shown, respectively,
in Fig. \ref{fig3} and \ref{fig4}. As for the spectrum of region 1 in 
Tol 1214--277, they are characterised by strong emission lines. 
However, the emission lines of the high ionization ions, except for the
He {\sc ii} $\lambda$4686 emission line, are not present in the spectra of 
these regions implying milder ionizing radiation as compared to Tol 1214--277.
The corrected fluxes $I(\lambda)/I({\rm H}\beta)$ and the equivalent widths 
$EW$ of emission lines together with the extinction coefficient $C$(H$\beta$), 
the absolute flux $F$(H$\beta$) of the H$\beta$ emission line and 
the equivalent width $EW$(abs) of hydrogen absorption lines 
for regions 1+2 and 3+4 of Tol 65 are listed in Table \ref{tab2}. 
Note that the flux of the He {\sc ii} $\lambda$4686 emission line relative 
to H$\beta$ is
stronger in the more evolved starbursts of regions 3+4, with lower equivalent 
width of H$\beta$, as compared to the younger bursts in regions 1+2.

We derive a redshift $z$ = 0.02603 $\pm$ 0.00008 for region 1 in 
Tol 1214--277 using the 44 brightest emission lines. As for Tol 65, 
redshifts $z$ = 0.00974 $\pm$ 0.00010 for
regions 1+2, using the 40 brightest lines, and $z$ = 0.00966 $\pm$ 0.00010 of
regions 3+4, using the 29 brightest lines, are obtained.

%*********************************************************
%*********************************************************
%  Tab.2 - emission line intensities in high resolution spectra of Tol 1214-277 and Tol 65
%*********************************************************
%*********************************************************

%\renewcommand{\baselinestretch}{1.0}
\begin{table*}%[h]%[tbh]
%     \centering{
\caption{Corrected
fluxes and equivalent widths of the emission lines in the high-resolution spectra 
of the brightest H {\sc ii} regions in Tol 1214--277 (slit \#3) and Tol 65
(slit \#6).}
\label{tab2}
\begin{tabular}{lcrccrccr} \hline \hline
  &\multicolumn{2}{c}{Tol 1214--277}&&\multicolumn{5}{c}{Tol 65} \\ \cline{2-3} 
\cline{5-9}
  &\multicolumn{2}{c}{region 1}&&\multicolumn{2}{c}{regions 1+2}&&\multicolumn{2}{c}{regions 3+4} \\ \cline{2-3}
\cline{5-6} \cline{8-9}
$\lambda_{0}$(\AA) Ion                  
&$I$($\lambda$)/$I$(H$\beta$)&$EW$ (\AA)&
&$I$($\lambda$)/$I$(H$\beta$)&$EW$ (\AA)&
&$I$($\lambda$)/$I$(H$\beta$)&$EW$ (\AA)
   \\ \hline
3704\ H16                      &  0.0153 $\pm$0.0018& 2.0 $\pm$0.2 &
                               &          ...       &      ...~~~~ &
                               &          ...       &      ...~~~~ \\
3712\ H15                      &  0.0160 $\pm$0.0017& 2.2 $\pm$0.1&
                               &          ...       &      ...~~~~ &
                               &          ...       &      ...~~~~ \\
3727\ [O {\sc ii}]             &  0.3584 $\pm$0.0061&47.8 $\pm$0.4&
                               &  0.5561 $\pm$0.0092&97.8 $\pm$0.4 &
                               &  0.7865 $\pm$0.0177&37.2 $\pm$1.3 \\
3750\ H12                      &  0.0328 $\pm$0.0018& 5.0 $\pm$0.2&
                               &  0.0574 $\pm$0.0027& 5.0 $\pm$0.2&
                               &  0.0454 $\pm$0.0298& 0.8 $\pm$0.6 \\
3770\ H11                      &  0.0419 $\pm$0.0018& 6.2 $\pm$0.2&
                               &  0.0660 $\pm$0.0026& 6.1 $\pm$0.2&
                               &  0.0514 $\pm$0.0262& 1.1 $\pm$0.6 \\
3798\ H10                      &  0.0590 $\pm$0.0020& 8.9 $\pm$0.2&
                               &  0.0797 $\pm$0.0025& 8.2 $\pm$0.2&
                               &  0.0675 $\pm$0.0201& 1.9 $\pm$0.6 \\
3820\ He {\sc i}               &  0.0132 $\pm$0.0015& 2.0 $\pm$0.2&
                               &  0.0102 $\pm$0.0012& 2.0 $\pm$0.2&
                               &          ...       &      ...~~~~ \\
3835\ H9                       &  0.0639 $\pm$0.0020& 9.9 $\pm$0.2&
                               &  0.0946 $\pm$0.0025&11.5 $\pm$0.2&
                               &  0.0749 $\pm$0.0166& 2.1 $\pm$0.6 \\
3868\ [Ne {\sc iii}]           &  0.3739 $\pm$0.0062&57.5 $\pm$0.4&
                               &  0.2816 $\pm$0.0047&48.8 $\pm$0.3&
                               &  0.2932 $\pm$0.0088&12.8 $\pm$0.7 \\
3889\ H8\ +\ He {\sc i}        &  0.2182 $\pm$0.0040&35.1 $\pm$0.3&
                               &  0.2125 $\pm$0.0039&35.5 $\pm$0.3&
                               &  0.2058 $\pm$0.0147& 8.4 $\pm$0.6 \\
3968\ [Ne {\sc iii}]\ +\ H7    &  0.3192 $\pm$0.0054&54.2 $\pm$0.4&
                               &  0.2790 $\pm$0.0049&42.0 $\pm$0.3&
                               &  0.2553 $\pm$0.0155&10.3 $\pm$0.6 \\
4026\ He {\sc i}               &  0.0196 $\pm$0.0010& 3.4 $\pm$0.2&
                               &  0.0158 $\pm$0.0010& 2.9 $\pm$0.2&
                               &  0.0132 $\pm$0.0053& 0.6 $\pm$0.4 \\
4068\ [S {\sc ii}]             &  0.0051 $\pm$0.0010& 0.9 $\pm$0.2&
                               &  0.0095 $\pm$0.0010& 1.6 $\pm$0.2&
                               &          ...       &      ...~~~~ \\
4101\ H$\delta$                &  0.2814 $\pm$0.0047&52.6 $\pm$0.4&
                               &  0.2719 $\pm$0.0046&48.8 $\pm$0.3&
                               &  0.2691 $\pm$0.0134&12.7 $\pm$0.6 \\
4227\ [Fe {\sc v}]             &  0.0068 $\pm$0.0008& 1.4 $\pm$0.2&
                               &        ...      &         ...~~~~ &
                               &        ...      &         ...~~~~  \\
4340\ H$\gamma$                &  0.4981 $\pm$0.0078&111.5 $\pm$0.6&
                               &  0.4661 $\pm$0.0073&96.8 $\pm$0.5&
                               &  0.4714 $\pm$0.0134&27.3 $\pm$0.7 \\
4363\ [O {\sc iii}]            &  0.1775 $\pm$0.0031&40.5 $\pm$0.4&
                               &  0.0975 $\pm$0.0020&22.7 $\pm$0.3&
                               &  0.0941 $\pm$0.0044& 5.5 $\pm$0.6 \\
4389\ He {\sc i}               &  0.0049 $\pm$0.0009& 1.2 $\pm$0.2&
                               &  0.0036 $\pm$0.0009& 1.0 $\pm$0.2&
                               &        ...      &         ...~~~~  \\
4471\ He {\sc i}               &  0.0364 $\pm$0.0011& 9.2 $\pm$0.3&
                               &  0.0352 $\pm$0.0011& 8.7 $\pm$0.2&
                               &  0.0312 $\pm$0.0034& 2.0 $\pm$0.4 \\
4658\ [Fe {\sc iii}]           &  0.0039 $\pm$0.0008& 1.1 $\pm$0.2&
                               &  0.0052 $\pm$0.0007& 1.4 $\pm$0.2&
                               &  0.0072 $\pm$0.0027& 0.5 $\pm$0.4 \\
4686\ He {\sc ii}              &  0.0531 $\pm$0.0013&15.2 $\pm$0.3&
                               &  0.0077 $\pm$0.0009& 2.2 $\pm$0.2&
                               &  0.0233 $\pm$0.0029& 1.5 $\pm$0.5 \\
4711\ [Ar {\sc iv}]+He {\sc i} &  0.0294 $\pm$0.0010& 8.6 $\pm$0.3&
                               &  0.0123 $\pm$0.0008& 3.4 $\pm$0.2&
                               &  0.0079 $\pm$0.0034& 0.5 $\pm$0.3 \\
4725\ [Ne {\sc iv}]            &  0.0020 $\pm$0.0004& 0.6 $\pm$0.2&
                               &         ...        &     ...~~~~ &
                               &         ...        &     ...~~~~  \\
4740\ [Ar {\sc iv}]            &  0.0187 $\pm$0.0009& 5.6 $\pm$0.3&
                               &  0.0064 $\pm$0.0006& 1.8 $\pm$0.2&
                               &  0.0066 $\pm$0.0052& 0.5 $\pm$0.3 \\
4861\ H$\beta$                 &  1.0000 $\pm$0.0149&311.6 $\pm$1.0&
                               &  1.0000 $\pm$0.0150&267.7 $\pm$0.8&
                               &  1.0000 $\pm$0.0183& 77.0 $\pm$1.0 \\
%4881\ [Fe {\sc iii}]           &         ...        &     ...~~~~ &
%                               &  0.0027 $\pm$0.0005& 0.8 $\pm$0.2&
%                               &         ...        &     ...~~~~ \\
4922\ He {\sc i}               &  0.0093 $\pm$0.0008& 3.1 $\pm$0.2&
                               &  0.0093 $\pm$0.0007& 2.9 $\pm$0.2&
                               &          ...       &     ...~~~~ \\
4959\ [O {\sc iii}]            &  1.7625 $\pm$0.0259&570.4 $\pm$1.3&
                               &  1.2645 $\pm$0.0188&355.7 $\pm$0.9&
                               &  1.1589 $\pm$0.0190& 91.1 $\pm$1.0 \\
4988\ [Fe {\sc iii}]           &  0.0046 $\pm$0.0005& 1.4 $\pm$0.2&
                               &  0.0048 $\pm$0.0005& 1.4 $\pm$0.2&
                               &          ...       &     ...~~~~ \\
5007\ [O {\sc iii}]            &  5.2819 $\pm$0.0770&1696.0 $\pm$2.2&
                               &  3.7565 $\pm$0.0555&1054.2 $\pm$1.4&
                               &  3.4617 $\pm$0.0548& 279.0 $\pm$1.6 \\
5146\ [Fe {\sc vi}]            &  0.0015 $\pm$0.0006& 0.6 $\pm$0.2&
                               &        ...     &         ...~~~~ &
                               &        ...     &         ...~~~~     \\
5177\ [Fe {\sc vi}]            &  0.0008 $\pm$0.0005& 0.3 $\pm$0.2&
                               &        ...     &         ...~~~~ &
                               &        ...     &         ...~~~~     \\
5200\ [N {\sc i}]              &  0.0010 $\pm$0.0006& 0.4 $\pm$0.2&
                               &  0.0035 $\pm$0.0007& 1.3 $\pm$0.2&
                               &        ...     &        ...~~~~  \\
5271\ [Fe {\sc iii}]           &  0.0014 $\pm$0.0009& 0.6 $\pm$0.4&
                               &  0.0027 $\pm$0.0009& 1.0 $\pm$0.3&
                               &        ...     &        ...~~~~  \\
5411\ He {\sc ii}              &  0.0037 $\pm$0.0006& 1.6 $\pm$0.3&
                               &        ...     &         ...~~~~ &
                               &        ...     &         ...~~~~     \\
5517\ [Cl {\sc iii}]           &  0.0022 $\pm$0.0005& 1.0 $\pm$0.3&
                               &  0.0019 $\pm$0.0004& 0.8 $\pm$0.2&
                               &        ...     &         ...~~~~     \\
5537\ [Cl {\sc iii}]           &  0.0011 $\pm$0.0005& 0.5 $\pm$0.2&
                               &  0.0016 $\pm$0.0004& 0.7 $\pm$0.2&
                               &        ...     &         ...~~~~     \\
5721\ [Fe {\sc vii}]           &  0.0012 $\pm$0.0005& 0.6 $\pm$0.2&
                               &        ...      &         ...~~~~ &
                               &        ...      &         ...~~~~  \\
5876\ He {\sc i}               &  0.0927 $\pm$0.0018&50.4 $\pm$0.6&
                               &  0.1043 $\pm$0.0019&48.5 $\pm$0.5&
                               &  0.0966 $\pm$0.0035&12.7 $\pm$0.8 \\
6087\ [Fe {\sc vii}]           &  0.0023 $\pm$0.0006& 1.3 $\pm$0.3&
                               &        ...      &         ...~~~~ &
                               &        ...      &         ...~~~~  \\
6300\ [O {\sc i}]              &  0.0080 $\pm$0.0005& 5.1 $\pm$0.3&
                               &  0.0178 $\pm$0.0006&10.1 $\pm$0.3&
                               &  0.0190 $\pm$0.0026& 2.9 $\pm$0.5 \\
6312\ [S {\sc iii}]            &  0.0076 $\pm$0.0005& 4.8 $\pm$0.3&
                               &  0.0114 $\pm$0.0005& 6.6 $\pm$0.3&
                               &  0.0149 $\pm$0.0018& 2.3 $\pm$0.5 \\
6363\ [O {\sc i}]              &  0.0032 $\pm$0.0005& 2.0 $\pm$0.3&
                               &  0.0062 $\pm$0.0006& 3.7 $\pm$0.7&
                               &        ...     &         ...~~~~     \\
6548\ [N {\sc ii}]             &  0.0035 $\pm$0.0005& 1.8 $\pm$0.3&
                               &  0.0061 $\pm$0.0005& 2.3 $\pm$0.2&
                               &        ...     &         ...~~~~     \\
6563\ H$\alpha$                &  2.7389 $\pm$0.0435&1798.1 $\pm$3.3&
                               &  2.7667 $\pm$0.0444&1601.8 $\pm$2.5&
                               &  2.6651 $\pm$0.0458& 422.4 $\pm$2.5 \\
6583\ [N {\sc ii}]             &  0.0094 $\pm$0.0006& 4.7 $\pm$0.3&
                               &  0.0171 $\pm$0.0006& 9.3 $\pm$0.3&
                               &  0.0214 $\pm$0.0018& 4.1 $\pm$0.7 \\
6678\ He {\sc i}               &  0.0249 $\pm$0.0008&18.6 $\pm$0.5&
                               &  0.0269 $\pm$0.0007&17.4 $\pm$0.4&
                               &  0.0256 $\pm$0.0017& 4.2 $\pm$0.7 \\
6717\ [S {\sc ii}]             &  0.0192 $\pm$0.0006&14.4 $\pm$0.5&
                               &  0.0503 $\pm$0.0011&32.9 $\pm$0.5&
                               &  0.0646 $\pm$0.0022&10.8 $\pm$0.8 \\
6731\ [S {\sc ii}]             &  0.0159 $\pm$0.0006&12.0 $\pm$0.4&
                               &  0.0402 $\pm$0.0010&26.3 $\pm$0.4&
                               &  0.0439 $\pm$0.0019& 7.4 $\pm$0.8 \\
%7005\ [Ar {\sc v}]             &  0.0006 $\pm$0.0004& 0.5 $\pm$0.2&
%                               &        ...      &         ...~~~~ &
%                               &        ...      &         ...~~~~  \\
7065\ He {\sc i}               &  0.0256 $\pm$0.0008&22.5 $\pm$0.6&
                               &  0.0296 $\pm$0.0008&23.0 $\pm$0.5&
                               &  0.0238 $\pm$0.0018& 4.6 $\pm$0.9 \\
7135\ [Ar {\sc iii}]           &  0.0215 $\pm$0.0007&19.0 $\pm$0.6&
                               &  0.0281 $\pm$0.0008&23.3 $\pm$0.5&
                               &  0.0315 $\pm$0.0020& 6.6 $\pm$0.9 \\
7281\ He {\sc i}               &  0.0050 $\pm$0.0005& 4.6 $\pm$0.4&
                               &  0.0066 $\pm$0.0005& 6.1 $\pm$0.5&
                               &  0.0044 $\pm$0.0017& 1.0 $\pm$0.7 \\
7320\ [O {\sc ii}]             &  0.0051 $\pm$0.0005& 4.7 $\pm$0.4&
                               &  0.0115 $\pm$0.0005&11.2 $\pm$0.4&
                               &  0.0146 $\pm$0.0017& 3.2 $\pm$0.9 \\
7330\ [O {\sc ii}]             &  0.0045 $\pm$0.0004& 4.2 $\pm$0.4&
                               &  0.0099 $\pm$0.0005&10.0 $\pm$0.4&
                               &  0.0124 $\pm$0.0015& 2.7 $\pm$0.9 \\
%                     & & & & & \\
$C$(H$\beta$)\ dex             &\multicolumn {2}{c}{0.105 $\pm$0.019}&&\multicolumn {2}{c}{0.065 $\pm$0.019}&
&\multicolumn {2}{c}{0.000 $\pm$0.020}  \\
$F$(H$\beta$)$^{\rm a}$              &\multicolumn {2}{c}{1.62 $\pm$0.01}  &&\multicolumn {2}{c}{1.64 $\pm$0.01}&
&\multicolumn {2}{c}{0.38 $\pm$0.01}    \\
%$EW$(H$\beta$)~\AA             &\multicolumn {2}{c}{241 $\pm$1}      &&\multicolumn {2}{c}{250 $\pm$2}&&\multicolumn {2}{c}{77 $\pm$1}        \\
$EW$(abs)~\AA                  &\multicolumn {2}{c}{0.0 $\pm$0.2}    &&\multicolumn {2}{c}{5.9 $\pm$0.2}&
&\multicolumn {2}{c}{1.3 $\pm$0.6}      \\
\hline%\hline
\end{tabular}

$^{\rm a}$in units 10$^{-14}$\ erg\ s$^{-1}$cm$^{-2}$.
%}
\end{table*}
%\renewcommand{\baselinestretch}{1.5}

%*********************************************************
%*********************************************************

%*********************************************************
%*********************************************************
%  Tab.3 - emission line intensities in low resolution spectra of Tol 1214-277
%*********************************************************
%*********************************************************

%\renewcommand{\baselinestretch}{1.0}
\begin{table*}%[h]%[tbh]
%     \centering{
\caption{Corrected
fluxes and equivalent widths of the emission lines in the low-resolution spectra 
of the brightest H {\sc ii} regions in Tol 1214--277 (slit \#1).}
\label{tab3}
\begin{tabular}{lcrccrccrccr} \hline \hline
  &\multicolumn{5}{c}{Region 1}&&\multicolumn{5}{c}{Region 2} \\ \cline{2-6} 
\cline{8-12}
  &\multicolumn{2}{c}{1999}&&\multicolumn{2}{c}{2000}&
&\multicolumn{2}{c}{1999}&&\multicolumn{2}{c}{2000} \\ \cline{2-3}
\cline{5-6} \cline{8-9} \cline{11-12}
$\lambda_{0}$(\AA) Ion                  
&$I$($\lambda$)/$I$(H$\beta$)&$EW$$^{\rm a}$&
&$I$($\lambda$)/$I$(H$\beta$)&$EW$$^{\rm a}$&
&$I$($\lambda$)/$I$(H$\beta$)&$EW$$^{\rm a}$&
&$I$($\lambda$)/$I$(H$\beta$)&$EW$$^{\rm a}$
   \\ \hline
3727\ [O {\sc ii}]             &  0.237$\pm$0.007&40.6&
                               &  0.285$\pm$0.007&42.6 &
                               &  1.928$\pm$0.342&53.8 &
                               &  1.361$\pm$0.209&40.8 \\
3750\ H12                      &  0.044$\pm$0.009& 4.0&
                               &  0.034$\pm$0.006& 4.3&
                               &         ...      & ... &
                               &         ...      & ... \\
3770\ H11                      &  0.050$\pm$0.008& 5.1&
                               &  0.049$\pm$0.006& 6.7&
                               &         ...      & ... &
                               &         ...      & ... \\
3798\ H10                      &  0.064$\pm$0.007& 7.5&
                               &  0.067$\pm$0.006& 9.5&
                               &         ...      & ... &
                               &         ...      & ... \\
3820\ He {\sc i}               &  0.009$\pm$0.003& 1.5&
                               &  0.009$\pm$0.002& 1.4&
                               &         ...      & ... &
                               &         ...      & ... \\
3835\ H9                       &  0.068$\pm$0.003& 8.2&
                               &  0.061$\pm$0.005& 8.7&
                               &         ...      & ... &
                               &         ...      & ... \\
3868\ [Ne {\sc iii}]           &  0.321$\pm$0.006&57.4&
                               &  0.357$\pm$0.007&56.4&
                               &         ...      & ... &
                               &         ...      & ... \\
3889\ H8\ +\ He {\sc i}        &  0.208$\pm$0.006&35.3&
                               &  0.216$\pm$0.006&34.0&
                               &         ...      & ... &
                               &         ...      & ... \\
3968\ [Ne {\sc iii}]\ +\ H7    &  0.297$\pm$0.007&53.6&
                               &  0.317$\pm$0.007&53.5&
                               &         ...      & ... &
                               &         ...      & ... \\
4026\ He {\sc i}               &  0.017$\pm$0.002& 3.4&
                               &  0.017$\pm$0.002& 3.1&
                               &         ...      & ... &
                               &         ...      & ... \\
4068\ [S {\sc ii}]             &  0.006$\pm$0.001& 1.3&
                               &  0.009$\pm$0.002& 1.6&
                               &         ...      & ... &
                               &         ...      & ... \\
4101\ H$\delta$                &  0.263$\pm$0.006&52.1&
                               &  0.277$\pm$0.006&53.1&
                               &         ...      & ... &
                               &         ...      & ... \\
4227\ [Fe {\sc v}]             &  0.010$\pm$0.002& 2.3&
                               &  0.009$\pm$0.002& 1.8 &
                               &         ...      & ... &
                               &         ...      & ... \\
4340\ H$\gamma$                &  0.476$\pm$0.008&107.5&
                               &  0.488$\pm$0.008&110.0&
                               &  0.418$\pm$0.093& 7.3&
                               &  0.482$\pm$0.081&15.6 \\
4363\ [O {\sc iii}]            &  0.171$\pm$0.003&40.6&
                               &  0.172$\pm$0.003&39.8&
                               &         ...      & ... &
                               &         ...      & ... \\
4389\ He {\sc i}               &  0.005$\pm$0.001& 1.3&
                               &  0.005$\pm$0.001& 1.2&
                               &         ...      & ... &
                               &         ...      & ... \\
4471\ He {\sc i}               &  0.033$\pm$0.002& 8.3&
                               &  0.037$\pm$0.001&10.0&
                               &         ...      & ... &
                               &         ...      & ... \\
4658\ [Fe {\sc iii}]           &         ...      & ... &
                               &  0.004$\pm$0.001& 1.1&
                               &         ...      & ... &
                               &         ...      & ... \\
4686\ He {\sc ii}              &  0.057$\pm$0.002&16.6&
                               &  0.053$\pm$0.001&15.0&
                               &         ...      & ... &
                               &         ...      & ... \\
4711\ [Ar {\sc iv}]+He {\sc i} &  0.033$\pm$0.001& 9.5&
                               &  0.032$\pm$0.001& 9.3&
                               &         ...      & ... &
                               &         ...      & ... \\
4740\ [Ar {\sc iv}]            &  0.023$\pm$0.001& 6.7&
                               &  0.020$\pm$0.001& 6.0&
                               &         ...      & ... &
                               &         ...      & ... \\
4861\ H$\beta$                 &  1.000$\pm$0.015&320.2&
                               &  1.000$\pm$0.015&315.3&
                               &  1.000$\pm$0.086& 32.6&
                               &  1.000$\pm$0.068& 50.2 \\
4922\ He {\sc i}               &  0.011$\pm$0.001& 3.7&
                               &  0.008$\pm$0.001& 2.6&
                               &         ...      & ... &
                               &         ...      & ... \\
4959\ [O {\sc iii}]            &  1.710$\pm$0.025&580.1&
                               &  1.736$\pm$0.025&571.7&
                               &  0.493$\pm$0.052& 20.9&
                               &  0.380$\pm$0.035& 16.4 \\
5007\ [O {\sc iii}]            &  5.127$\pm$0.075&1794.0&
                               &  5.220$\pm$0.075&1789.0&
                               &  1.447$\pm$0.102& 63.7&
                               &  1.314$\pm$0.073& 59.7 \\
5200\ [N {\sc i}]              &  0.003$\pm$0.001& 1.4&
                               &         ...      & ... &
                               &         ...      & ... &
                               &         ...      & ... \\
%5271\ [Fe {\sc iii}]           &         ...      & ... &
%                               &         ...      & ... &
%                               &         ...      & ... &
%                               &         ...      & ... \\
5411\ He {\sc ii}              &  0.004$\pm$0.001& 1.6&
                               &  0.004$\pm$0.001& 1.7&
                               &         ...      & ... &
                               &         ...      &  ... \\
%5517\ [Cl {\sc iii}]           &         ...      &  ... &
%                               &         ...      &  ... &
%                               &         ...      & ... &
%                               &         ...      &  ... \\
%5537\ [Cl {\sc iii}]           &         ...      &  ... &
%                               &         ...      &  ... &
%                               &         ...      & ... &
%                               &         ...      &  ... \\
5876\ He {\sc i}               &  0.090$\pm$0.002&50.5&
                               &  0.095$\pm$0.002&49.9&
                               &         ...      & ... &
                               &         ...      & ... \\
6300\ [O {\sc i}]              &  0.010$\pm$0.001& 6.5&
                               &  0.009$\pm$0.001& 5.7&
                               &         ...      & ... &
                               &         ...      & ... \\
%6312\ [S {\sc iii}]            &         ...      & ... &
%                               &         ...      & ... &
%                               &         ...      & ... &
%                               &         ...      & ... \\
6363\ [O {\sc i}]              &  0.003$\pm$0.001& 2.2&
                               &  0.003$\pm$0.001& 2.0&
                               &         ...      & ... &
                               &         ...      & ... \\
6563\ H$\alpha$                &  2.401$\pm$0.038&1799.0&
                               &  2.739$\pm$0.043&1807.0&
                               &  2.879$\pm$0.196&227.2&
                               &  2.878$\pm$0.153&268.1 \\
6583\ [N {\sc ii}]             &  0.007$\pm$0.001& 3.7&
                               &  0.010$\pm$0.001& 5.7&
                               &         ...      & ... &
                               &         ...      & ... \\
6678\ He {\sc i}               &  0.025$\pm$0.001&21.4&
                               &  0.026$\pm$0.001&19.3&
                               &         ...      & ... &
                               &         ...      & ... \\
6717\ [S {\sc ii}]             &  0.016$\pm$0.001&13.8&
                               &  0.018$\pm$0.001&13.2&
                               &         ...      & ... &
                               &         ...      & ... \\
6731\ [S {\sc ii}]             &  0.014$\pm$0.001&12.3&
                               &  0.015$\pm$0.001&10.6&
                               &         ...      & ... &
                               &         ...      & ... \\
7065\ He {\sc i}               &  0.027$\pm$0.001&22.6&
                               &  0.026$\pm$0.001&22.4&
                               &         ...      & ... &
                               &         ...      & ... \\
7135\ [Ar {\sc iii}]           &  0.022$\pm$0.001&20.1&
                               &  0.022$\pm$0.001&19.1&
                               &         ...      & ... &
                               &         ...      & ... \\
%7281\ He {\sc i}               &         ...      & ... &
%                               &         ...      & ... &
%                               &         ...      & ... &
%                               &         ...      & ... \\
%7320\ [O {\sc ii}]             &         ...      & ... &
%                               &         ...      & ... &
%                               &         ...      & ... &
%                               &         ...      & ... \\
%7330\ [O {\sc ii}]             &         ...      & ... &
%                               &         ...      & ... &
%                               &         ...      & ... &
%                               &         ...      & ... \\
                     & & & & & \\
$C$(H$\beta$)\ dex             &\multicolumn {2}{c}{0.085$\pm$0.019}&&\multicolumn {2}{c}{0.065$\pm$0.019}&
&\multicolumn {2}{c}{0.000$\pm$0.073}&&\multicolumn {2}{c}{0.030$\pm$0.063}  \\
$F$(H$\beta$)$^{\rm b}$              &\multicolumn {2}{c}{1.69$\pm$0.01}  &&\multicolumn {2}{c}{1.62$\pm$0.01}&
&\multicolumn {2}{c}{0.02$\pm$0.01}&&\multicolumn {2}{c}{0.02$\pm$0.01}    \\
%$EW$(H$\beta$)~\AA             &\multicolumn {2}{c}{241$\pm$1}      &&\multicolumn {2}{c}{250$\pm$2}&&\multicolumn {2}{c}{77$\pm$1}        \\
$EW$(abs)~\AA                  &\multicolumn {2}{c}{3.5$\pm$0.7}    &&\multicolumn {2}{c}{0.9$\pm$0.6}&
&\multicolumn {2}{c}{3.9$\pm$0.6}&&\multicolumn {2}{c}{0.1$\pm$1.0}      \\
\hline%\hline
\end{tabular}

$^{\rm a}$in \AA.

$^{\rm b}$in units 10$^{-14}$\ erg\ s$^{-1}$cm$^{-2}$.
%}
\end{table*}
%\renewcommand{\baselinestretch}{1.5}

%*********************************************************
%*********************************************************

%*********************************************************
%*********************************************************
%  Tab.4 - emission line intensities in low resolution spectra of Tol 65
%*********************************************************
%*********************************************************

%\renewcommand{\baselinestretch}{1.0}
\begin{table*}%[h]%[tbh]
%     \centering{
\caption{Corrected
fluxes and equivalent widths of the emission lines in the low-resolution spectra 
of the brightest H {\sc ii} regions in Tol 65 (slit \#4).}
\label{tab4}
\begin{tabular}{lcrccrccrccr} \hline \hline
  &\multicolumn{2}{c}{Region 1}&&\multicolumn{2}{c}{Region 2}&&\multicolumn{2}{c}{Region 3}&&\multicolumn{2}{c}{Region 4} \\ 
\cline{2-3} \cline{5-6} \cline{8-9} \cline{11-12} 
$\lambda_{0}$(\AA) Ion                  
&$I$($\lambda$)/$I$(H$\beta$)&$EW$$^{\rm a}$&
&$I$($\lambda$)/$I$(H$\beta$)&$EW$$^{\rm a}$&
&$I$($\lambda$)/$I$(H$\beta$)&$EW$$^{\rm a}$&
&$I$($\lambda$)/$I$(H$\beta$)&$EW$$^{\rm a}$
   \\ \hline
3727\ [O {\sc ii}]             &  0.449$\pm$0.010&91.6&
                               &  0.684$\pm$0.021&78.2 &
                               &  0.739$\pm$0.080&18.4 &
                               &  0.614$\pm$0.030&33.7 \\
3750\ H12                      &  0.030$\pm$0.005& 6.0&
                               &  0.019$\pm$0.012& 2.2&
                               &         ...     & ... &
                               &         ...     & ... \\
3770\ H11                      &  0.038$\pm$0.005& 7.7&
                               &  0.022$\pm$0.013& 2.5&
                               &         ...     & ... &
                               &         ...     & ... \\
3798\ H10                      &  0.044$\pm$0.005& 8.5&
                               &  0.037$\pm$0.013& 4.3&
                               &         ...     & ... &
                               &         ...     & ... \\
%3820\ He {\sc i}               &         ...     & ... &
%                               &         ...     & ... &
%                               &         ...     & ... &
%                               &         ...     & ... \\
3835\ H9                       &  0.062$\pm$0.005&12.1&
                               &  0.059$\pm$0.014& 6.5&
                               &         ...     & ... &
                               &         ...     & ... \\
3868\ [Ne {\sc iii}]           &  0.313$\pm$0.006&60.7&
                               &  0.250$\pm$0.010&26.9&
                               &  0.278$\pm$0.046& 5.4 &
                               &  0.304$\pm$0.015&14.0 \\
3889\ H8\ +\ He {\sc i}        &  0.180$\pm$0.006&33.8&
                               &  0.177$\pm$0.014&19.0&
                               &  0.189$\pm$0.084& 1.9 &
                               &  0.153$\pm$0.016& 7.1 \\
3968\ [Ne {\sc iii}]\ +\ H7    &  0.275$\pm$0.007&51.0&
                               &  0.237$\pm$0.013&25.7&
                               &  0.226$\pm$0.063& 2.5 &
                               &  0.249$\pm$0.017&11.8 \\
4026\ He {\sc i}               &  0.016$\pm$0.002& 3.0&
                               &         ...     & ... &
                               &         ...     & ... &
                               &         ...     & ... \\
4068\ [S {\sc ii}]             &  0.009$\pm$0.001& 1.8&
                               &         ...     & ... &
                               &         ...     & ... &
                               &         ...     & ... \\
4101\ H$\delta$                &  0.253$\pm$0.006&51.6&
                               &  0.247$\pm$0.012&29.1&
                               &  0.243$\pm$0.052& 3.6 &
                               &  0.229$\pm$0.014&11.9 \\
4340\ H$\gamma$                &  0.493$\pm$0.009&111.6&
                               &  0.485$\pm$0.013&65.2&
                               &  0.485$\pm$0.036&10.1&
                               &  0.511$\pm$0.014&32.1 \\
4363\ [O {\sc iii}]            &  0.122$\pm$0.003&28.0&
                               &  0.094$\pm$0.005&12.8&
                               &  0.097$\pm$0.029& 2.3 &
                               &  0.119$\pm$0.008& 7.4 \\
4389\ He {\sc i}               &  0.005$\pm$0.002& 1.3&
                               &         ...     & ... &
                               &         ...     & ... &
                               &         ...     & ... \\
4471\ He {\sc i}               &  0.036$\pm$0.001& 9.1&
                               &  0.038$\pm$0.004& 5.9&
                               &         ...     & ... &
                               &  0.022$\pm$0.006& 1.5 \\
4658\ [Fe {\sc iii}]           &  0.004$\pm$0.001& 1.1 &
                               &         ...     & ... &
                               &         ...     & ... &
                               &         ...     & ... \\
4686\ He {\sc ii}              &  0.006$\pm$0.001& 1.8&
                               &         ...     & ... &
                               &         ...     & ... &
                               &  0.034$\pm$0.005& 2.6 \\
4711\ [Ar {\sc iv}]+He {\sc i} &  0.014$\pm$0.001& 4.1&
                               &  0.012$\pm$0.002& 2.1&
                               &         ...     & ... &
                               &         ...     & ... \\
4740\ [Ar {\sc iv}]            &  0.007$\pm$0.001& 2.1&
                               &         ...     & ... &
                               &         ...     & ... &
                               &         ...     & ... \\
4861\ H$\beta$                 &  1.000$\pm$0.015&310.9&
                               &  1.000$\pm$0.017&191.3&
                               &  1.000$\pm$0.032& 33.9&
                               &  1.000$\pm$0.019& 89.3 \\
4922\ He {\sc i}               &  0.007$\pm$0.001& 2.3&
                               &         ...     & ... &
                               &         ...     & ... &
                               &         ...     & ... \\
4959\ [O {\sc iii}]            &  1.341$\pm$0.020&430.5&
                               &  1.095$\pm$0.018&217.5&
                               &  1.034$\pm$0.030&37.8&
                               &  1.284$\pm$0.022&122.4 \\
5007\ [O {\sc iii}]            &  4.043$\pm$0.059&1371.0&
                               &  3.277$\pm$0.051&659.0&
                               &  3.136$\pm$0.076&119.4&
                               &  3.810$\pm$0.062&363.3 \\
5200\ [N {\sc i}]              &  0.005$\pm$0.001& 1.8&
                               &         ...     & ... &
                               &         ...     & ... &
                               &         ...     & ... \\
5271\ [Fe {\sc iii}]           &  0.003$\pm$0.001& 1.4&
                               &         ...     & ... &
                               &         ...     & ... &
                               &         ...     & ... \\
5517\ [Cl {\sc iii}]           &  0.003$\pm$0.001& 1.3 &
                               &         ...     & ... &
                               &         ...     & ... &
                               &         ...     & ... \\
5537\ [Cl {\sc iii}]           &  0.002$\pm$0.001& 1.0 &
                               &         ...     & ... &
                               &         ...     & ... &
                               &         ...     & ... \\
5876\ He {\sc i}               &  0.104$\pm$0.002&54.8&
                               &  0.099$\pm$0.003&34.2&
                               &  0.092$\pm$0.011& 5.5 &
                               &  0.096$\pm$0.004&14.8 \\
6300\ [O {\sc i}]              &  0.020$\pm$0.001&12.7&
                               &  0.024$\pm$0.002& 9.4&
                               &         ...     & ... &
                               &         ...     & ... \\
%6312\ [S {\sc iii}]            &         ...      & ... &
%                               &         ...      & ... &
%                               &         ...      & ... &
%                               &         ...      & ... \\
6363\ [O {\sc i}]              &  0.006$\pm$0.001&12.7&
                               &         ...     & ... &
                               &         ...     & ... &
                               &         ...     & ... \\
6563\ H$\alpha$                &  2.749$\pm$0.043&1807.0&
                               &  2.748$\pm$0.046&1173.0&
                               &  2.451$\pm$0.065&194.9&
                               &  2.526$\pm$0.045&560.8 \\
6583\ [N {\sc ii}]             &  0.013$\pm$0.001& 7.0&
                               &  0.024$\pm$0.004&12.1&
                               &         ...     & ... &
                               &         ...     & ... \\
6678\ He {\sc i}               &  0.028$\pm$0.001&19.6&
                               &  0.029$\pm$0.003&13.0&
                               &         ...     & ... &
                               &  0.031$\pm$0.004& 7.5 \\
6717\ [S {\sc ii}]             &  0.043$\pm$0.001&30.5&
                               &  0.072$\pm$0.002&31.2&
                               &  0.071$\pm$0.009& 6.1 &
                               &         ...     & ... \\
6731\ [S {\sc ii}]             &  0.033$\pm$0.001&23.8&
                               &  0.044$\pm$0.002&19.4&
                               &  0.042$\pm$0.007& 3.7 &
                               &         ...     & ... \\
7065\ He {\sc i}               &  0.034$\pm$0.001&27.7&
                               &  0.027$\pm$0.002&14.9&
                               &         ...     & ... &
                               &  0.027$\pm$0.004& 7.7 \\
7135\ [Ar {\sc iii}]           &  0.028$\pm$0.001&24.2&
                               &  0.034$\pm$0.003&19.0&
                               &         ...     & ... &
                               &  0.033$\pm$0.004& 9.5 \\
7281\ He {\sc i}               &  0.006$\pm$0.001& 5.1 &
                               &  0.008$\pm$0.003& 4.9 &
                               &         ...     & ... &
                               &         ...     & ... \\
%7320\ [O {\sc ii}]             &         ...      & ... &
%                               &         ...      & ... &
%                               &         ...      & ... &
%                               &         ...      & ... \\
%7330\ [O {\sc ii}]             &         ...      & ... &
%                               &         ...      & ... &
%                               &         ...      & ... &
%                               &         ...      & ... \\
                     & & & & & \\
$C$(H$\beta$)\ dex             &\multicolumn {2}{c}{0.045$\pm$0.019}&&\multicolumn {2}{c}{0.000$\pm$0.020}&
&\multicolumn {2}{c}{0.000$\pm$0.030}&&\multicolumn {2}{c}{0.000$\pm$0.021}  \\
$F$(H$\beta$)$^{\rm b}$              &\multicolumn {2}{c}{1.79$\pm$0.01}  &&\multicolumn {2}{c}{0.54$\pm$0.01}&
&\multicolumn {2}{c}{0.12$\pm$0.01}&&\multicolumn {2}{c}{0.30$\pm$0.01}    \\
%$EW$(H$\beta$)~\AA             &\multicolumn {2}{c}{241$\pm$1}      &&\multicolumn {2}{c}{250$\pm$2}&&\multicolumn {2}{c}{77$\pm$1}        \\
$EW$(abs)~\AA                  &\multicolumn {2}{c}{0.3$\pm$0.8}    &&\multicolumn {2}{c}{0.0$\pm$1.2}&
&\multicolumn {2}{c}{1.8$\pm$0.4}&&\multicolumn {2}{c}{0.0$\pm$0.5}      \\
\hline%\hline
\end{tabular}

$^{\rm a}$in \AA.

$^{\rm b}$in units 10$^{-14}$\ erg\ s$^{-1}$cm$^{-2}$.
%}
\end{table*}
%\renewcommand{\baselinestretch}{1.5}

%*********************************************************
%*********************************************************
%  Tab.5 - element abundances from the high-resolution spectra
%*********************************************************
%*********************************************************

%\renewcommand{\baselinestretch}{1.0}
\begin{table*}%[tbh]
\caption{Physical conditions and element abundances derived from the high-resolution
spectra of the brightest H {\sc ii} regions of Tol 1214--277 (slit \#3) and 
Tol 65 (slit \#6).}
\label{tab5}
\begin{center}
\begin{tabular}{lccccc} \hline \hline
                                    & Tol 1214--277     &&\multicolumn{3}{c}{Tol 65} \\ \cline{2-2} \cline{4-6}
Value                               & region 1          && regions 1+2 && regions 3+4 \\ \hline
$T_e$(O {\sc iii})(K)                     &19970$\pm$250&&17270$\pm$210&&17720$\pm$460 \\
$T_e$(O {\sc ii})(K)                      &15680$\pm$180&&14750$\pm$170&&14920$\pm$370 \\
$T_e$(S {\sc iii})(K)                     &18280$\pm$210&&16030$\pm$170&&16400$\pm$390 \\
$N_e$(S {\sc ii})(cm$^{-3}$)              &   250$\pm$90&&   190$\pm$50&&   10$\pm$10  \\
$N_e$(He {\sc ii})(cm$^{-3}$)             &    70$\pm$40&&   260$\pm$50&&   20$\pm$10 \\
$\tau$($\lambda$3889)               &       0.0   &&      0.0   &&       0.0    \\ \\
O$^+$/H$^+$($\times$10$^5$)         &0.280$\pm$0.009&&0.517$\pm$0.018&&0.692$\pm$0.048\\
O$^{+2}$/H$^+$($\times$10$^5$)      &3.041$\pm$0.093&&2.943$\pm$0.093&&2.560$\pm$0.163\\
O$^{+3}$/H$^+$($\times$10$^6$)      &2.136$\pm$0.105&&0.305$\pm$0.036&&0.904$\pm$0.136 \\
O/H($\times$10$^5$)                 &3.534$\pm$0.094&&3.490$\pm$0.095&&3.343$\pm$0.170\\
12 + log(O/H)                       &7.548$\pm$0.012&&7.543$\pm$0.012&&7.524$\pm$0.022\\ \\
N$^{+}$/H$^+$($\times$10$^7$)       &0.644$\pm$0.032&&1.312$\pm$0.046&&1.603$\pm$0.127\\
ICF(N)$^a$                          &12.6\,~~~~~~~~~~&&6.76\,~~~~~~~~~~&&4.83\,~~~~~~~~~~\\
log(N/O)                            &--1.638$\pm$0.025~~&&--1.595$\pm$0.019~~&&--1.635$\pm$0.041~~\\ \\
Ne$^{+2}$/H$^+$($\times$10$^5$)     &0.436$\pm$0.014&&0.467$\pm$0.015&&0.455$\pm$0.031\\
ICF(Ne)$^a$                         &1.16\,~~~~~~~~~~&&1.19\,~~~~~~~~~~&&1.31\,~~~~~~~~~~\\
log(Ne/O)                           &--0.843$\pm$0.018~~&&--0.800$\pm$0.019~~&&--0.750$\pm$0.037~~\\ \\
S$^+$/H$^+$($\times$10$^7$)         &0.330$\pm$0.011&&0.937$\pm$0.023&&1.079$\pm$0.049\\
S$^{+2}$/H$^+$($\times$10$^7$)      &2.236$\pm$0.164&&4.772$\pm$0.257&&5.850$\pm$0.794\\
ICF(S)$^a$                          &2.91\,~~~~~~~~~~&&1.88\,~~~~~~~~~~&&1.56\,~~~~~~~~~~\\
log(S/O)                            &--1.676$\pm$0.030~~&&--1.513$\pm$0.023~~&&--1.490$\pm$0.055~~\\ \\
Cl$^{+2}$/H$^+$($\times$10$^8$)     &0.505$\pm$0.115&&0.691$\pm$0.121&&      ...   \\
ICF(Cl)$^a$                         &1.85\,~~~~~~~~~~&&1.77\,~~~~~~~~~~&&    ...   \\
log(Cl/O)                           &--3.578$\pm$0.100~~&&--3.455$\pm$0.077~~&& ... \\ \\
Ar$^{+2}$/H$^+$($\times$10$^7$)     &0.567$\pm$0.020&&0.908$\pm$0.028&&0.979$\pm$0.068\\
Ar$^{+3}$/H$^+$($\times$10$^7$)     &1.469$\pm$0.077&&0.671$\pm$0.067&&0.657$\pm$0.526\\
ICF(Ar)$^a$                         &1.01\,~~~~~~~~~~&&1.02\,~~~~~~~~~~&&1.04\,~~~~~~~~~~\\
log(Ar/O)                           &--2.237$\pm$0.021~~&&--2.337$\pm$0.023~~&&--2.295$\pm$0.143~~\\ \\
Fe$^{+2}$/H$^+$($\times$10$^7$)     &0.661$\pm$0.129&&1.027$\pm$0.149&&1.382$\pm$0.517\\
ICF(Fe)$^a$                         &15.8\,~~~~~~~~~~&&8.44\,~~~~~~~~~~&&6.04\,~~~~~~~~~~\\
log(Fe/O)                           &--1.530$\pm$0.086~~&&--1.605$\pm$0.064~~&&--1.603$\pm$0.164~~\\
$[$O/Fe$]$                          &0.110$\pm$0.086&&0.185$\pm$0.064&&0.183$\pm$0.164\\ \\
He$^+$/H$^+$($\lambda$4471)         &0.0777$\pm$0.0027&&0.0717$\pm$0.0023&&0.0671$\pm$0.0072\\
He$^+$/H$^+$($\lambda$5876)         &0.0769$\pm$0.0021&&0.0798$\pm$0.0018&&0.0803$\pm$0.0029\\
He$^+$/H$^+$($\lambda$6678)         &0.0755$\pm$0.0024&&0.0774$\pm$0.0021&&0.0760$\pm$0.0051\\
He$^+$/H$^+$(mean)                  &0.0767$\pm$0.0014&&0.0768$\pm$0.0012&&0.0779$\pm$0.0024\\
He$^{+2}$/H$^+$($\lambda$4686)      &0.0049$\pm$0.0001&&0.0007$\pm$0.0001&&0.0021$\pm$0.0003\\
He/H                                &0.0816$\pm$0.0014&&0.0775$\pm$0.0012&&0.0800$\pm$0.0024\\
$Y$                                 &0.2459$\pm$0.0043&&0.2365$\pm$0.0037&&0.2423$\pm$0.0075\\ \hline
\end{tabular}
\end{center}

~~~~~~~~~~~~~~~~~~~~~~~~~~~~~~~~$^a$ICF is the ionization correction factor.
\end{table*}
%\renewcommand{\baselinestretch}{1.5}

%*********************************************************
%*********************************************************

%*********************************************************
%*********************************************************
%  Tab.6 - element abundances from the low-resolution spectra
%*********************************************************
%*********************************************************

%\renewcommand{\baselinestretch}{1.0}
\begin{table*}%[tbh]
\caption{Physical conditions and element abundances derived from the low-resolution
spectra of the brightest H {\sc ii} regions of Tol 1214--277 (slit \#1) and 
Tol 65 (slit \#4).}
\label{tab6}
\begin{tabular}{lccccccc} \hline \hline
                                    & \multicolumn{2}{c}{Tol 1214--277 (region 1)}&&\multicolumn{4}{c}{Tol 65} \\ \cline{2-3} \cline{5-8}
Value                               &1999            &2000            && region 1       & region 2       & region 3       & region 4 \\ \hline
$T_e$(O {\sc iii})(K)               &19850$\pm$260   &19760$\pm$250   &&18740$\pm$250   &18240$\pm$500   &19030$\pm$3260  &19100$\pm$740   \\
$T_e$(O {\sc ii})(K)                &15640$\pm$190   &15650$\pm$180   &&15300$\pm$190   &15110$\pm$390   &15390$\pm$2460  &15410$\pm$560   \\
$T_e$(S {\sc iii})(K)               &18170$\pm$220   &18100$\pm$210   &&17260$\pm$210   &16840$\pm$420   &17500$\pm$2700  &17550$\pm$610   \\
$N_e$(S {\sc ii})(cm$^{-3}$)        &  410$\pm$150   &  220$\pm$90    &&   140$\pm$60   &    10$\pm$10   &   10$\pm$10    &   10$\pm$10    \\
$N_e$(He {\sc ii})(cm$^{-3}$)       &    10$\pm$1    &    70$\pm$1    &&   220$\pm$50   &    10$\pm$1    &       ...      &    10$\pm$3    \\
$\tau$($\lambda$3889)               &       0.3      &       0.0      &&      0.2       &      0.2       &       ...      &      0.2       \\ \\
O$^+$/H$^+$($\times$10$^5$)         &0.189$\pm$0.008 &0.224$\pm$0.009 &&0.372$\pm$0.014 &0.579$\pm$0.043 &0.594$\pm$0.250 &0.491$\pm$0.051 \\
O$^{+2}$/H$^+$($\times$10$^5$)      &2.989$\pm$0.096 &3.067$\pm$0.095 &&2.645$\pm$0.087 &2.276$\pm$0.150 &1.985$\pm$0.784 &2.407$\pm$0.217 \\
O$^{+3}$/H$^+$($\times$10$^6$)      &2.181$\pm$0.119 &2.019$\pm$0.100 &&0.214$\pm$0.031 &      ...       &       ...      &0.947$\pm$0.201 \\
O/H($\times$10$^5$)                 &3.397$\pm$0.097 &3.494$\pm$0.096 &&3.039$\pm$0.088 &2.855$\pm$0.156 &2.578$\pm$0.823 &2.993$\pm$0.224 \\
12 + log(O/H)                       &7.531$\pm$0.012 &7.543$\pm$0.012 &&7.483$\pm$0.013 &7.456$\pm$0.024 &7.411$\pm$0.139 &7.476$\pm$0.032 \\ \\
N$^{+}$/H$^+$($\times$10$^7$)       &0.497$\pm$0.038 &0.672$\pm$0.048 &&0.894$\pm$0.040 &1.786$\pm$0.213 &       ...      &       ...      \\
ICF(N)$^{\rm a}$                    &17.9\,~~~~~~~~~~&15.6\,~~~~~~~~~~&&8.16\,~~~~~~~~~~&4.93\,~~~~~~~~~~&       ...      &       ...      \\
log(N/O)                            &--1.581$\pm$0.035~~&--1.524$\pm$0.033~~&&--1.620$\pm$0.023~~&--1.511$\pm$0.057~~&   ...   &   ...     \\ \\
Ne$^{+2}$/H$^+$($\times$10$^5$)     &0.380$\pm$0.013 &0.426$\pm$0.014 &&0.424$\pm$0.015 &0.361$\pm$0.026 &0.362$\pm$0.147&0.393$\pm$0.038  \\
ICF(Ne)$^{\rm a}$                   &1.14\,~~~~~~~~~~&1.14\,~~~~~~~~~~&&1.15\,~~~~~~~~~~&1.25\,~~~~~~~~~~&1.30\,~~~~~~~~~~&1.24\,~~~~~~~~~~\\
log(Ne/O)                           &--0.895$\pm$0.019~~&--0.857$\pm$0.019~~&&--0.796$\pm$0.020~~&--0.800$\pm$0.040~~&--0.739$\pm$0.225~~&--0.787$\pm$0.053~~\\ \\
%S$^+$/H$^+$($\times$10$^7$)         &0.330$\pm$0.011 &0.306$\pm$0.010 &&0.740$\pm$0.019 &1.131$\pm$0.052 &1.070$\pm$0.265 &       ...      \\
%S$^{+2}$/H$^+$($\times$10$^7$)      &      ...       &      ...       &&      ...       &      ...       &       ...      &       ...      \\
%ICF(S)$^{\rm a}$                    &      ...       &      ...       &&      ...       &     ...        &       ...      &       ...      \\
%log(S/O)                            &      ...         &      ...         &&  ...           &  ...           &   ...      &       ...      \\ \\
Cl$^{+2}$/H$^+$($\times$10$^8$)     &      ...       &      ...       &&0.857$\pm$0.144 &      ...       &       ...      &       ...      \\
ICF(Cl)$^{\rm a}$                   &      ...       &      ...       &&1.91\,~~~~~~~~~~&      ...       &       ...      &       ...      \\
log(Cl/O)                           &      ...         &      ...         &&--3.270$\pm$0.074~~&      ...       & ...     &       ...      \\ \\
Ar$^{+2}$/H$^+$($\times$10$^7$)     &0.595$\pm$0.027 &0.583$\pm$0.020 &&0.798$\pm$0.029 &1.007$\pm$0.081 &       ...      &0.933$\pm$0.116 \\
Ar$^{+3}$/H$^+$($\times$10$^7$)     &1.789$\pm$0.109 &1.618$\pm$0.086 &&0.622$\pm$0.095 &      ...       &       ...      &       ...      \\
ICF(Ar)$^{\rm a}$                   &1.01\,~~~~~~~~~~&1.01\,~~~~~~~~~~&&1.01\,~~~~~~~~~~&1.89\,~~~~~~~~~~&       ...      &2.11\,~~~~~~~~~~\\
log(Ar/O)                           &--2.152$\pm$0.023~~&--2.198$\pm$0.021~~&&--2.325$\pm$0.033~~&--2.176$\pm$0.042~~&  ... &--2.182$\pm$0.063~~\\ \\
Fe$^{+2}$/H$^+$($\times$10$^7$)     &      ...       &0.692$\pm$0.012 &&0.728$\pm$0.184 &      ...       &       ...      &       ...      \\
ICF(Fe)$^{\rm a}$                   &      ...       &19.5\,~~~~~~~~~~&&10.2\,~~~~~~~~~~&      ...       &       ...      &       ...      \\
log(Fe/O)                           &      ...       &--1.381$\pm$0.073~~&&--1.639$\pm$0.110~~&      ...       & ...      &       ...      \\
$[$O/Fe$]$                          &      ...       &--0.006$\pm$0.074~~&&0.192$\pm$0.111&      ...       &     ...      &       ...      \\ \\
He$^+$/H$^+$($\lambda$4471)         &0.0726$\pm$0.0032&0.0783$\pm$0.0031&&0.0742$\pm$0.0029&0.0816$\pm$0.0086&     ...    &0.0481$\pm$0.0119\\
He$^+$/H$^+$($\lambda$5876)         &0.0768$\pm$0.0015&0.0788$\pm$0.0014&&0.0805$\pm$0.0015&0.0827$\pm$0.0026&     ...    &0.0810$\pm$0.0035\\
He$^+$/H$^+$($\lambda$6678)         &0.0768$\pm$0.0027&0.0788$\pm$0.0024&&0.0806$\pm$0.0026&0.0854$\pm$0.0076&     ...    &0.0943$\pm$0.0122\\
He$^+$/H$^+$(mean)                  &0.0762$\pm$0.0012&0.0787$\pm$0.0011&&0.0795$\pm$0.0012&0.0829$\pm$0.0023&     ...    &0.0795$\pm$0.0033\\
He$^{+2}$/H$^+$($\lambda$4686)      &0.0053$\pm$0.0001&0.0049$\pm$0.0001&&0.0006$\pm$0.0001&        ...      &     ...    &0.0031$\pm$0.0005\\
He/H                                &0.0815$\pm$0.0012&0.0836$\pm$0.0012&&0.0801$\pm$0.0012&0.0829$\pm$0.0023&     ...    &0.0826$\pm$0.0033\\
$Y$                                 &0.2457$\pm$0.0038&0.2504$\pm$0.0036&&0.2425$\pm$0.0036&0.2488$\pm$0.0072&     ...    &0.2482$\pm$0.0102\\ \hline
\end{tabular}

$^{\rm a}$ICF is the ionization correction factor.
\end{table*}

\subsubsection{Low-resolution spectra}

We show in Fig. \ref{fig5} the low-resolution spectra of regions 1 and 2 
in Tol 1214--277 obtained during the two 1999 and 2000 runs, with slit \#1 
(Fig. \ref{fig1}) oriented along the major axis of the galaxy. 
The corrected fluxes and 
equivalent widths of the emission lines are shown in Table \ref{tab3}.
As in the case of the high-resolution spectra, the high-ionization 
He {\sc ii} $\lambda$4686, $\lambda$5411 and [Fe {\sc v}] $\lambda$4227 
emission
lines were also detected in region 1. The low spectral resolution
precludes flux measurement of the [S {\sc iii}] $\lambda$6312 emission
line, blended with the [O {\sc i}] $\lambda$6300 emission line. Likewise,
the [N {\sc ii}] $\lambda$6583 emission line is blended with H$\alpha$.
The [N {\sc ii}] $\lambda$6583 fluxes in the spectra of region 1 were 
therefore measured by setting
the continuum at the top of the wing of the H$\alpha$ emission line. 
Region 2 is characterised by a blue spectrum, however only a few strongest 
emission lines with low equivalent widths are detected. The 
temperature-sensitive 
[O {\sc iii}] $\lambda$4363 emission line is not seen, precluding 
a reliable abundance determination.
%Region 2 is likely the region on the post-starburst stage.

We show the four spectra of regions 1 to 4 in Tol 65 
in Fig. \ref{fig6}. The corrected fluxes and
equivalent widths of emission lines are shown in Table \ref{tab4}. Note
that the observed H$\alpha$/H$\beta$ flux ratios in regions 3 and 4 are
lower than the theoretical recombination value. This likely
reflects uncertainties in the flux calibration of the Tol 65 spectrum.
Therefore, for regions 3 and 4 we set the extinction coefficient 
$C$(H$\beta$) = 0. The high spatial resolution allows to identify 
more accurately the regions with He {\sc ii}
$\lambda$4686 emission. It is seen from Table \ref{tab4} that 
the He {\sc ii} $\lambda$4686/H$\beta$ flux ratio is small in region 1 and 
much larger in region 4. No He {\sc ii} emission is
detected in regions 2 and 3. 
Hence, the ionization radiation is the hardest in region 4.

%
% --------------------------------------
\subsection{Heavy element abundances}
% --------------------------------------

To derive element abundances, we adopted a two-zone photoionized H {\sc ii} 
region model (Stasi\'nska \cite{S90}) including a high-ionization zone 
with temperature $T_{\rm e}$(O {\sc iii}), and a low-ionization zone with 
temperature $T_{\rm e}$(O {\sc ii}).  
The electron temperature $T_{\rm e}$(O {\sc iii}) is derived from the 
[O {\sc iii}]$\lambda$4363/($\lambda$4959+$\lambda$5007) ratio 
using a five-level atom model. That temperature is used for the 
derivation of the He$^{+}$, He$^{+2}$, O$^{+2}$, Ne$^{+2}$ and Ar$^{+3}$ ionic 
abundances. 
$T_{\rm e}$(O {\sc ii}) is obtained from $T_{\rm e}$(O {\sc iii}) using
the relation given in Izotov et al. (\cite{ITL94}), 
based on a fit to the photoionization models of Stasi\'nska (\cite{S90}). The 
temperature $T_{\rm e}$(O {\sc ii}) is 
used to derive the O$^+$, N$^+$, S$^+$ 
and Fe$^{+2}$ ion abundances. For S$^{+2}$, Cl$^{+2}$ and Ar$^{+2}$
we have adopted an electron temperature intermediate between
$T_{\rm e}$(O {\sc iii}) and $T_{\rm e}$(O {\sc ii}) following the 
prescriptions of Garnett (\cite{G92}).

The oxygen abundance is derived as
\begin{equation}
{\rm \frac{O}{H} = \frac{O^+}{H^+} + \frac{O^{+2}}{H^+} + \frac{O^{+3}}{H^+}},
\label{eq:O} 
\end{equation}
where
\begin{equation}
{\rm \frac{O^{+3}}{H^+} = \frac{He^{+2}}{He^+}\times\left(\frac{O^+}{H^+} + 
\frac{O^{+2}}{H^+}\right)}.
\end{equation}

Total abundances of other heavy elements were computed after correction for 
unseen stages of ionization, as described in Izotov et al. (\cite{ITL94}), 
Thuan et al. (\cite{TIL95}) and Guseva et al. (\cite{G03}).

%*********************************************************
%*********************************************************

%*********************************************************
%*********************************************************

\subsubsection{High-resolution spectra}

The heavy element abundances obtained from the high-resolution spectra of 
region 1 in Tol 1214--277 (slit \#3) and of regions 1+2 and 3+4 in Tol 65 
(slit \#6) are shown in 
Table \ref{tab5}. We also show in this Table the electron temperatures 
$T_{\rm e}$(O {\sc iii}), $T_{\rm e}$(O {\sc ii}),
and $T_{\rm e}$(S {\sc iii}), the
electron number density $N_{\rm e}$(S {\sc ii}) and $N_{\rm e}$(He {\sc ii}), 
and the ionization correction
factors ICF used for the abundance determination. 

The oxygen abundances of Tol 1214--277 and Tol 65
are very similar. They are in general agreement with those obtained 
in previous studies.

The high spectral resolution allows us to determine the nitrogen abundance
with great accuracy because the [N {\sc ii}] $\lambda$6583 emission line
is not blended with H$\alpha$. The
nitrogen-to-oxygen abundance ratios log N/O = --1.60 - --1.64 obtained here
agree well with those found by Izotov et al. (\cite{I01a}), but differ 
significantly from the value obtained by Pagel et al. (\cite{P92}). 
For Tol 1214--277,
Pagel et al. (\cite{P92}) derived a higher value, log N/O = --1.46 $\pm$ 0.06, 
while, for
Tol 65, they found log N/O = --1.81 $\pm$ 0.15, lower than 
the mean for the most metal-deficient BCDs (Izotov \& Thuan 
\cite{IT99}). Our new determination of the N/O abundance ratio 
in these two BCDs supports
the finding by Thuan et al. (\cite{TIL95}) and Izotov \& Thuan (\cite{IT99})
that log N/O for BCDs with 12 + log O/H $\la$ 7.60 
lies in a narrow range, with a mean value of --1.60.

\begin{figure}[h]%[hbtp]
    \psfig{figure=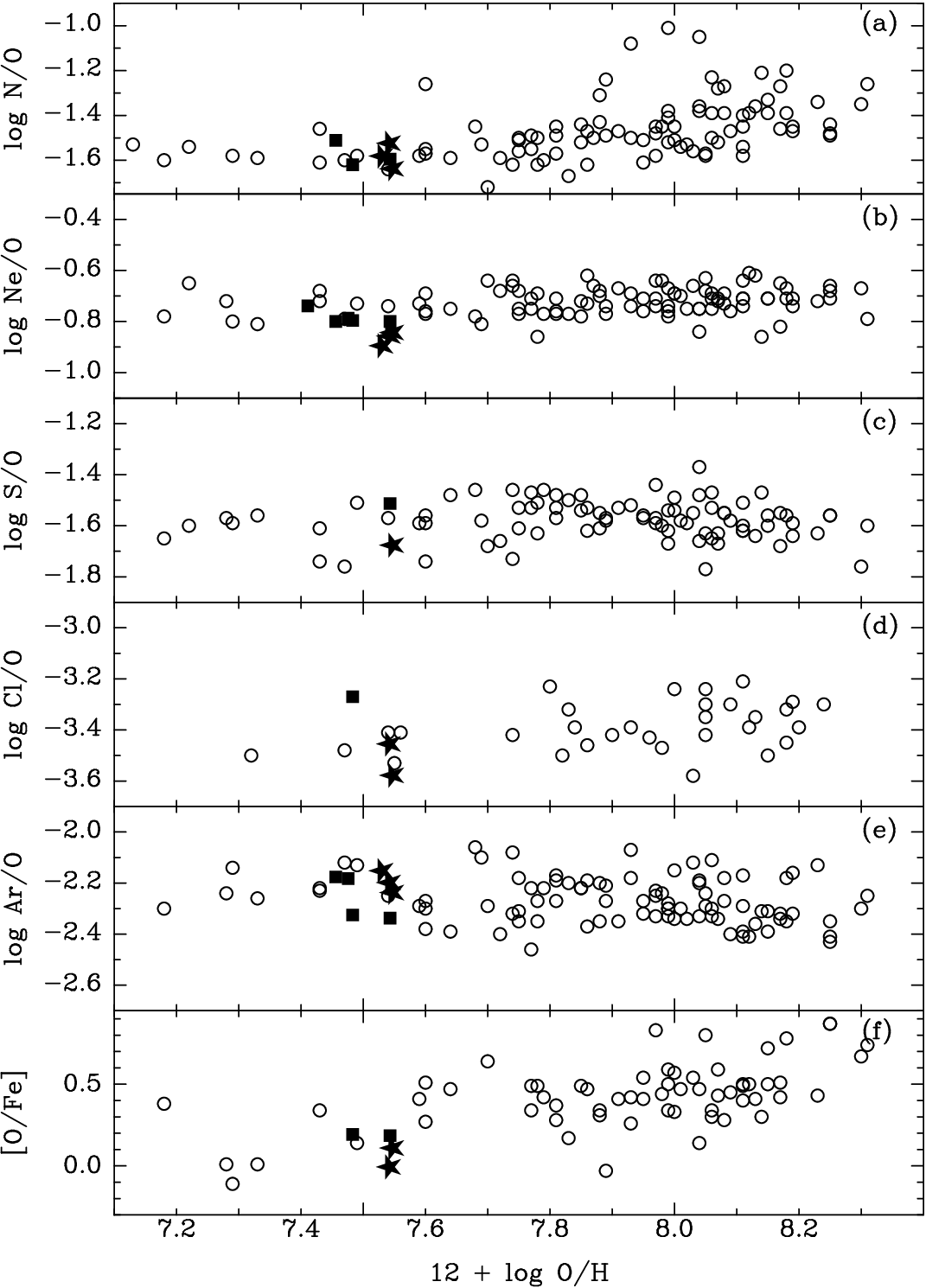,angle=0,width=7.5cm}
    \caption{Abundance ratios in Tol 1214--277 (stars) and Tol 65 (squares)
as a function of oxygen abundance 12 + log O/H: 
(a) N/O; (b) Ne/O; (c) S/O; (d) Cl/O; (e) Ar/O; (f) [O/Fe]
$\equiv$ log(O/Fe) -- log(O/Fe)$_\odot$, where log(O/Fe)$_\odot$ = 
--1.42 (Anders \& Grevesse \cite{Anders89}). Open circles are the data
from Izotov \& Thuan (\cite{IT99,IT04}), Lipovetsky et al. (\cite{L99}) and
Guseva et al. (\cite{G03}).}
    \label{fig7}
\end{figure}

In Fig. \ref{fig7} we compare several heavy element-to-oxygen abundance 
ratios determined from the high- and low-resolution spectra for Tol 1214--277 
(filled stars) and Tol 65 (filled squares) with those derived 
for a sample of 93 BCDs by Izotov \& Thuan (\cite{IT99,IT04}), 
Lipovetsky et al. (\cite{L99})
and Guseva et al. (\cite{G03}) (open circles). The heavy
element abundance ratios in Tol 1214--277 and Tol 65 are in general 
agreement with those derived for other very low-metallicity BCDs.

Note the increasing trend of [O/Fe] with increasing 12 + log O/H 
(open symbols in Fig. \ref{fig7}f). The relatively low [O/Fe] in Tol 1214--277 and Tol 65 (filled symbols in Fig. \ref{fig7}f) further supports that
trend. Such a trend can be explained by a larger
Fe depletion onto dust in galaxies with larger oxygen abundance.
%, assuming
%that the non-depleted Fe/O abundance ratio is independent on metallicity
%(Izotov et al. \cite{I04b}).

\subsubsection{Low-resolution spectra}

The heavy element abundances obtained from the low-resolution spectra, 
together with the adopted electron temperatures, electron number densities
and ionization correction factors, are shown in Table \ref{tab6} for region 1
in Tol 1214--277, observed in 1999 and 2000 at slit position \#1, and 
regions 1 to 4 in Tol 65, observed in 1999 at slit position \#4.
The oxygen abundance for region 1 in Tol 1214--277 derived from the 
low-resolution observations is very similar to that obtained from the
high-resolution observations (Table \ref{tab5}). The other heavy element
abundances are also in good agreement. The oxygen abundance for regions 1 -- 4
in Tol 65 derived from the low-resolution spectra is slightly lower than that 
found for regions 1+2 and 3+4 from the high-resolution spectra. The agreement
between other heavy element abundances is good.

%
% ------------------------------
\subsection{Helium abundance}
% ------------------------------
%
He emission-line strengths in the high- and low-resolution spectra
of region 1 in Tol 1214--277 and regions 1--4 in Tol 65 are converted to 
singly ionized helium $y^+$ $\equiv$ He$^+$/H$^+$ and doubly ionized helium 
$y^{+2}$ $\equiv$ He$^{+2}$/H$^+$
abundances, using the theoretical He {\sc i} recombination line emissivities 
by Smits (\cite{S96}). 

%*********************************************************
%*********************************************************
%  Fig.7 - the distribution of the heavy elements
%*********************************************************
%*********************************************************

%*********************************************************
%*********************************************************

%                    Table 7
%*********************************************************
% General characteristics of the field galaxies
%*********************************************************

\begin{table}%[hbtp]
%     \centering{
\caption{Coordinates of the emission-line galaxies in the fields of 
Tol 1214--277 and Tol 65.}
\label{tab7}
\begin{tabular}{lcc} \hline \hline
Name         & $\alpha$(J2000.0) &$\delta$(J2000.0) \\ \hline %\hline %& $V$   & $V-I$  \\ \hline \hline
Tol 1214--277$^a$&12$^{\rm h}$17$^{\rm m}$17\fs1         &--28$^\circ$02\arcmin33\arcsec     \\ %  & 17.30 & --0.35 \\
{\it g}1               &12~\,17\,~\,17.0         &--28 02 27~~     \\ %  & 17.30 & --0.35 \\
{\it g}2               &12~\,17\,~\,15.5         &--28 02 49~~     \\ %  & 17.30 & --0.35 \\
{\it g}3               &12~\,17\,~\,14.0         &--28 02 53~~     \\ %  & 17.30 & --0.35 \\
{\it g}4               &12~\,17\,~\,25.0         &--28 00 17~~     \\ \\ %  & 17.30 & --0.35 \\ 
Tol 65$^a$       &12~\,25\,~\,46.9         &--36 14 01~~     \\ %  & 17.30 & --0.35 \\
{\it g}5               &12~\,25\,~\,45.6         &--36 14 13~~     \\ %  & 17.30 & --0.35 \\
{\it g}6               &12~\,25\,~\,51.5         &--36 13 12~~     \\ %  & 17.30 & --0.35 \\
{\it g}7               &12~\,25\,~\,59.1         &--36 16 05~~     \\ %  & 17.30 & --0.35 \\
{\it g}8               &12~\,25\,~\,42.9         &--36 13 24~~     \\ \hline %  & 17.30 & --0.35 \\ \hline
\end{tabular}

$^a$ Coordinates of the brightest region 1.
\end{table}

The helium mass fraction was calculated as
\begin{equation}
Y=\frac{4y[1-20({\rm O/H})]}{1+4y},                     \label{eq:Y}
\end{equation}
where $y$ = $y^+$ + $y^{+2}$ is the number density of helium relative to 
hydrogen (Pagel et al. \cite{P92}).

The main mechanisms causing deviations of the 
observed He {\sc i} emission line 
fluxes from their theoretical values are collisional and fluorescent 
enhancement. In order to correct for these effects, we have adopted the 
procedure, discussed in more detail in Izotov et al. (\cite{ITL94,ITL97})
and Izotov \& Thuan (\cite{IT98}).
The singly ionized helium abundance $y^+$ 
and He mass fraction $Y$ is obtained for each of the 
three He {\sc i} $\lambda$4471, $\lambda$5876 and $\lambda$6678 lines.
We then derive the weighted mean $y^+$ of these three determinations, the
weight of each line being scaled to its intensity.
The obtained ionic and total He abundances ($y^+$, $y^{+2}$, $y$) 
and mass fractions ($Y$) 
in Tol 1214--277 and Tol 65 are shown in 
Tables \ref{tab5} and \ref{tab6}. In general, they are consistent with the
He abundances found by Izotov et al. (\cite{I01a}) from the Keck 
observations of both BCDs. They are also consistent with the primordial
helium mass fraction $Y_{\rm p}$ = 0.244 -- 0.245 obtained by Izotov \& Thuan
(\cite{IT98}) and Izotov et al. (\cite{I99}). 
An exception is the He mass fraction
$Y$ determined from the high-resolution spectrum
of regions 1+2 in Tol 65, which is lower (Table \ref{tab5}). 
%This is mainly 
%because of the low flux of the He {\sc i} $\lambda$4471 (Table \ref{tab5})
%which is most subject to underlying stellar He {\sc i} absorption. 
However, at the 95\% confidence level, 
it is consistent with other determinations of $Y$ in Tol 65.

%Some systematic effects such as the difference between the electron 
%temperatures
%$T_{\rm e}$(He$^+$) and $T_{\rm e}$(O$^{+2}$), the presence of the 
%underlying stellar He {\sc i} absorption lines, the collisional excitation of 
%hydrogen emission lines may change the derived values of $Y$ in Tol 1214--277 
%and
%Tol 65. However, as shown by Izotov \& Thuan (\cite{IT04}), these additional
%effects largely cancel each other. Therefore, we do not consider them in 
%present paper.

%            Table 8
%*********************************************************
%  Line intensities in the field galaxies
%*********************************************************

\begin{table}%[hbtp]
%     \centering{
\caption{Parameters of the emission lines in the 
spectra of galaxies in the fields of Tol 1214--277 and Tol 65.}
\label{tab8}
\begin{tabular}{lcrrc} \hline \hline
$\lambda_{0}$(\AA) Ion & $\lambda_{obs}$(\AA)   &$F$($\lambda$)$^a$
&$EW$(\AA) &$z^b$ \\ \hline %\hline
  \multicolumn{5}{c}{{\it g}1 (slit \#3)} \\ %\cline{1-5} 
3727\ [O {\sc ii}]      & 4585.2$\pm$0.1 &  12.5$\pm$0.7  & 52 $\pm$~\,3 & 0.2303  \\
4340\ H$\gamma$         & 5338.2$\pm$0.3 &  1.7$\pm$0.3   &  7 $\pm$~\,1 & 0.2299  \\
4861\ H$\beta$          & 5979.4$\pm$0.2 &  4.3$\pm$0.5   & 18 $\pm$~\,2 & 0.2299  \\
4959\ [O {\sc iii}]     & 6100.5$\pm$0.7 &  2.0$\pm$0.8  &  6 $\pm$~\,2 & 0.2302  \\ 
5007\ [O {\sc iii}]     & 6159.6$\pm$0.4 &  5.1$\pm$1.1  & 15 $\pm$~\,3 & 0.2302  \\
\hline %\hline
  \multicolumn{5}{c}{{\it g}2 (slit \#2)} \\ %\cline{1-5} 
5007\ [O {\sc iii}]     & 5137.0$\pm$0.9 &  1.9$\pm$0.5  & 20 $\pm$~\,5 & 0.0260  \\
6563\ H$\alpha$         & 6732.5$\pm$0.8 &  2.3$\pm$0.5  & 43 $\pm$10 & 0.0259  \\
\hline%\hline
  \multicolumn{5}{c}{{\it g}3 (slit \#2)} \\ %\cline{1-5} 
3727\ [O {\sc ii}]      & 5999.0$\pm$0.4 &  12.2$\pm$1.0  & 47 $\pm$~\,4 & 0.6096  \\
4861\ H$\beta$          & 7821.1$\pm$0.4 &  6.7$\pm$0.8   & 26 $\pm$~\,3 & 0.6088  \\
\hline %\hline
  \multicolumn{5}{c}{{\it g}4 (slit \#1)} \\ %\cline{1-5} 
3727\ [O {\sc ii}]      & 6421.8$\pm$1.0 &   8.4$\pm$1.7   & 34 $\pm$~\,7 & 0.7231  \\
\hline %\hline
  \multicolumn{5}{c}{{\it g}5 (slit \#5)} \\ %\cline{1-5} 
3727\ [O {\sc ii}]      & 4105.7$\pm$1.0 &  10.6$\pm$1.6  & 51 $\pm$~\,8 & 0.1016  \\
4861\ H$\beta$          & 5350.1$\pm$1.1 &  1.2$\pm$0.6   &  4 $\pm$~\,2 & 0.1005  \\
4959\ [O {\sc iii}]     & 5463.2$\pm$2.6 &  3.3$\pm$1.2   & 13 $\pm$~\,5 & 0.1017  \\ 
5007\ [O {\sc iii}]     & 5515.9$\pm$0.9 &  3.9$\pm$0.9   & 15 $\pm$~\,4 & 0.1017  \\
6563\ H$\alpha$         & 7228.5$\pm$0.4 &  9.9$\pm$1.0   &118 $\pm$12 & 0.1014   \\
\hline %\hline
  \multicolumn{5}{c}{{\it g}6 (slit \#5)} \\ %\cline{1-5} 
3727\ [O {\sc ii}]      & 4942.4$\pm$0.2 & 107.7$\pm$3.6  & 27 $\pm$~\,1 & 0.3261  \\
4861\ H$\beta$          & 6445.4$\pm$0.4 &  34.4$\pm$3.2   &  7 $\pm$~\,1 & 0.3258  \\
5007\ [O {\sc iii}]     & 6637.2$\pm$0.5 &  34.5$\pm$2.9   &  7 $\pm$~\,1 & 0.3256  \\
\hline %\hline
  \multicolumn{5}{c}{{\it g}7 (slit \#6)} \\ %\cline{1-5} 
3727\ [O {\sc ii}]      & 5304.7$\pm$0.7 &   6.0$\pm$0.9   & 65 $\pm$~\,12 & 0.4233  \\
4861\ H$\beta$          & 6917.4$\pm$0.8 &   1.9$\pm$0.6   & 47 $\pm$~\,22 & 0.4229  \\
4959\ [O {\sc iii}]     & 7056.0$\pm$0.8 &   1.6$\pm$0.8   & 50 $\pm$~\,17 & 0.4229  \\
5007\ [O {\sc iii}]     & 7124.0$\pm$0.4 &   3.6$\pm$0.7   &230 $\pm$~\,17 & 0.4229  \\
\hline %\hline
  \multicolumn{5}{c}{{\it g}8 (slit \#6)} \\ %\cline{1-5} 
3727\ [O {\sc ii}]      & 4762.0$\pm$0.5 &   4.7$\pm$0.7   & 52 $\pm$~\,9 & 0.2777  \\
4861\ H$\beta$          & 6210.1$\pm$1.9 &   1.0$\pm$0.6   & 18 $\pm$~\,11 & 0.2775  \\
4959\ [O {\sc iii}]     & 6334.0$\pm$1.5 &   1.2$\pm$0.6   & 28 $\pm$~\,14 & 0.2773  \\
5007\ [O {\sc iii}]     & 6395.2$\pm$0.7 &   2.4$\pm$0.8   & 33 $\pm$~\,10 & 0.2773  \\
\hline %\hline
\end{tabular}

$^a$in units 10$^{-17}$\ erg\ s$^{-1}$cm$^{-2}$.  \\
$^b$ redshift.
%}
\end{table}
%***************************************************************
%***************************************************************

%*********************************************************
%*********************************************************
%  Fig.8 - the spectra of the galaxies in the field
%*********************************************************
%*********************************************************

\begin{figure*}
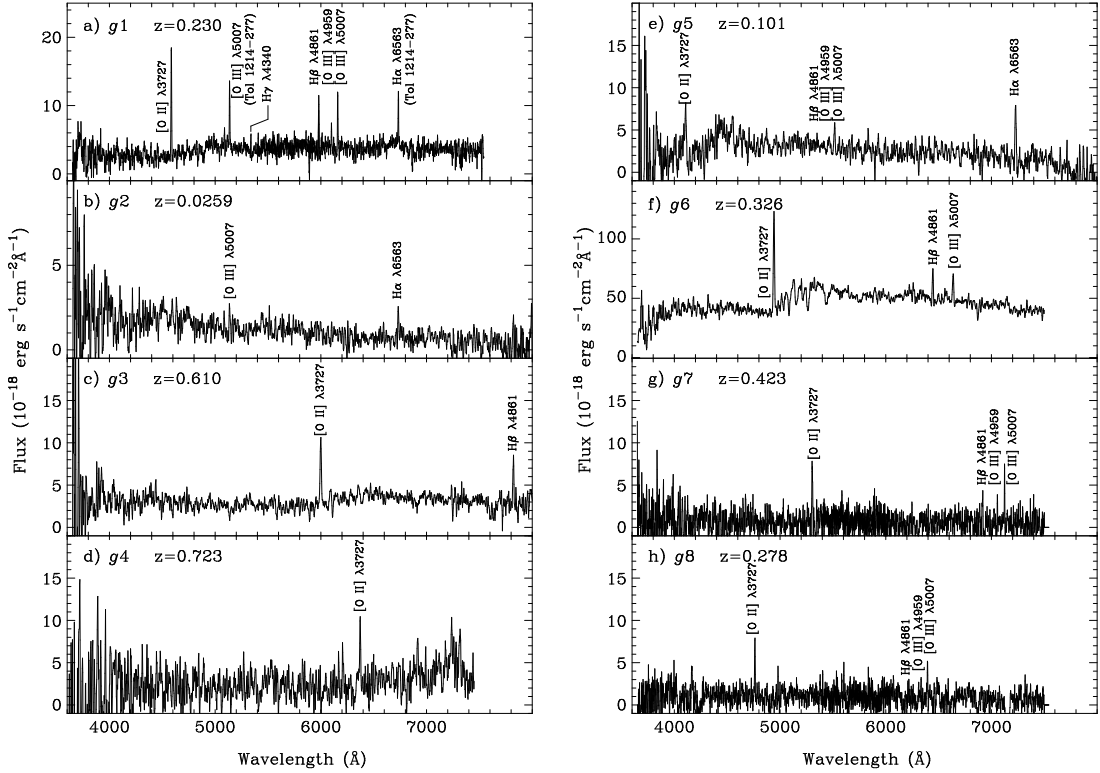
%[hbtp]
    \hspace*{1.5cm}\psfig{figure=0847fig8a.ps,angle=0,width=7.cm,clip=}
    \hspace*{0.4cm}\psfig{figure=0847fig8b.ps,angle=0,width=7.cm,clip=}
    \caption{Spectra of the emission-line galaxies in the fields of 
Tol 1214--277 (left panel) and Tol 65 (right panel).
The emission lines are labeled. The spectrum
of the galaxy {\it g}1 is contaminated by the emission lines of Tol 1214--277. All 
galaxies except for the galaxy {\it g}2 (b) are background galaxies. The galaxies 
Tol 1214--277 and {\it g}2 have the same redshift and are likely physically related.}
    \label{fig8}
\end{figure*}

%*********************************************************
%*********************************************************

%-----------------------------------------------------------------------------
\section{High-ionization emission lines in Tol 1214--277 \label{highion}}
%-----------------------------------------------------------------------------

Several emission lines of the high-ionization ions are detected in the
high-resolution spectrum of Tol 1214--277 (Table \ref{tab2}, Fig. \ref{fig2}).
The presence of these lines implies the existence of a 
substantial ionizing radiation field with
$\lambda$ $\la$ 228\AA\ (equivalent to photon energies $\ga$ 4 Ryd) 
in the brightest H {\sc ii} region 1.
In addition to confirming the findings of previous papers such as
[Fe {\sc v}] $\lambda$4227 and strong He {\sc ii} $\lambda$4686 emission, 
we detect here also 
emission lines of Ne {\sc iv} (the ionization potential of
Ne$^{+3}$ is 4.664 Ryd), Fe {\sc vi} (the ionization potential of
Fe$^{+5}$ is 5.513 Ryd) and Fe {\sc vii} (the ionization potential of
Fe$^{+6}$ is 7.281 Ryd). 
The large difference in the ionization potentials of these different
ions allows in principle to determine the slope of the hard radiation spectrum.

To constrain the nature of the hard ionizing radiation in Tol 1214--277, we
compare our observations with a  
spherically symmetric ionization-bounded H {\sc ii} 
region model calculated with the CLOUDY94 code (Ferland \cite{F96}; 
Ferland et al. \cite{F98}) for the
physical conditions of the brightest H {\sc ii} region. The input parameters
are the element abundances from Table \ref{tab5} and the flux of
ionizing photons $Q_{\rm H}$ (in s$^{-1}$) 
at $\lambda$912\AA\ as determined from the
H$\beta$ emission line luminosity $L$(H$\beta$). Because the spectrum of
Tol 1214--277 is not properly flux-calibrated, we adopt log $Q_{\rm H}$ = 52.7
from Izotov et al. (\cite{I04}). As source of ionizing
radiation, we chose to use the hottest 
CoStar stellar atmosphere model
F1, with a heavy element mass fraction $Z$ = 0.004 ($Z_\odot$/5) 
(the lowest $Z$ available for the CoStar models) and an effective temperature 
$T_{\rm eff}$ = 54000K (Schaerer \& de Koter \cite{S97}).
This model predicts harder radiation for
$\lambda$ $\la$ 228\AA\ as compared to Kurucz (\cite{K79})'s 
stellar atmosphere models.

The model predicts the He {\sc ii} $\lambda$4686 flux and the 
[Ne {\sc iv}] $\lambda$4725/[Ne {\sc iii}] $\lambda$3868 flux ratio to be  
respectively $\sim$ 50 and $\sim$ 200 times lower than the ones observed.
As for the high-ionization iron lines, the model predicts flux ratios
[Fe {\sc vi}] $\lambda$5146/[Fe {\sc iii}] $\lambda$4658 and
[Fe {\sc vii}] $\lambda$6087/[Fe {\sc iii}] $\lambda$4658 
respectively $\sim$ 500 and $\sim$ 10$^{5}$ times lower than the ones observed.
The disagreement between predicted and observed emission line fluxes
increases for ions of higher ionization degree. This implies that
the spectrum of hard radiation for energies $>$ 4 Ryd in Tol 1214--277 
increases more steeply at higher energies, as compared to the   
predictions of the CoStar models.
The disagreement between the observed and predicted fluxes and flux ratios 
is not likely
to diminish significantly if models with the lower Tol 1214--277 ionized gas
metallicity are used, because log $Q_{\rm He^+}$/$Q_{\rm H}$ is nearly 
constant for main-sequence stellar population models with 
$Z$ $\ga$ $Z_\odot$/50
(Schaerer \cite{S02,S03}). Here $Q_{\rm He^+}$ is the flux of ionizing
photons with $\lambda$ $\leq$ 228\AA\ ($\geq$ 4 Ryd).
In principle, it is possible to reproduce the 
observed He {\sc ii} $\lambda$4686
and [Ne {\sc iv}] $\lambda$4725 emission line fluxes by 
assuming that the massive
stars in the ionizing cluster of Tol 1214--277 have the very low metallicity,
of $Z$ $\la$ 10$^{-7}$. However, the predicted equivalent widths of
H$\beta$ and Ly$\alpha$ emission lines in such models 
(Schaerer \cite{S02,S03}) are several times
larger than those observed (Table \ref{tab2}; Thuan \& Izotov \cite{TI97}).
Furthermore, these models cannot reproduce the observed
fluxes of [Fe {\sc vi}] and [Fe {\sc vii}] emission lines. This suggests 
that
an additional compact source of X-ray emission is likely to be 
present in the
brightest H {\sc ii} region of Tol 1214--277. 
Fast shocks with velocities of $\sim$ 400 -- 500 km s$^{-1}$ 
(Dopita \& Sutherland \cite{DS96}) would reproduce
both the He {\sc ii} and Fe {\sc vii} emission.
High mass X-ray binary systems are also possible sources of X-ray emission. 

We now estimate the X-ray luminosity needed to account for the high-ionization
lines.
Using the value of $Q_{\rm H}$ derived
from the observations, we scale the spectral energy distribution
of stellar atmosphere model F1 to fit the line fluxes of the 
low-ionization ions. Adopting a multiplicative factor of 10$^5$, required 
to bring the ionizing radiation fluxes at 
$\lambda$ $\sim$ 100\AA\ 
($\sim$ 7 -- 8 Ryd) of the CoStar models to the level of  
the observed fluxes of the Fe {\sc vii} ions,
 we find 
that the X-ray luminosity of the ionizing cluster should be as high as
10$^{39}$ -- 10$^{40}$ erg s$^{-1}$, 
in agreement with the conclusions of 
Izotov et al. (\cite{I04}) who discovered unusually strong [Ne {\sc v}]
$\lambda$3426 emission in the spectrum of Tol 1214--277.
We note also that Thuan et al. (\cite{T04}) using {\sl Chandra} observations
have found in the two most metal-deficient BCDs known, I Zw 18 and 
SBS 0335--052, X-ray point sources with 0.5 -- 10 Kev luminosities in the 
range 1.3 -- 8.5 $\times$ 10$^{39}$ erg s$^{-1}$, just the range of 
X-ray luminosities predicted by our scaling argument.

\section{Emission-line galaxies in the fields of Tol 1214--277 and Tol 65
\label{field}}

One of the aims of this study was the search for emission-line companion
galaxies of Tol 1214--277 and Tol 65. For this we use spectroscopic
observations with different slit orientations (slits \#1 -- \#6 in
Table \ref{tab1} and Fig. \ref{fig1}), 
chosen to go through extended sources 
with irregular morphology and/or blue colour, as seen on deep VLT 
images of the fields around Tol 1214--277 and Tol 65 (Fricke et al. \cite{F01};
Papaderos et al. \cite{P99}).

We found in
total four emission-line galaxies in the field of Tol 1214--277 and four
emission-line galaxies in the field of Tol 65. These galaxies are labeled
in Fig. \ref{fig1}, and their equatorial coordinates at epoch J2000.0
are given in Table \ref{tab7}, along with the coordinates of 
Tol 1214--277 and Tol 65. Note that the galaxies {\it g}2 and 
{\it g}3 correspond to the galaxies $G_2$ and $G_3$ discussed by Fricke et al. (\cite{F01}), and
the galaxy {\it g}5 corresponds to the galaxy $G_1$ in Papaderos et al. (\cite{P99}).

The spectra of the emission-line galaxies in the fields of Tol 1214--277
and Tol 65 are shown respectively in the left and right panels of 
Fig. \ref{fig8}. In each spectrum, except for that of galaxy {\it g}4
(Fig. \ref{fig8}d), several emission lines are detected, most often the
 [O {\sc ii}] $\lambda$3727, H$\beta$ $\lambda$4861, [O {\sc iii}]
$\lambda$4959, $\lambda$5007 lines. In the spectrum of {\it g}4,
 only one emission
line is present. We identify the line to be [O {\sc ii}] $\lambda$3727, as
the continuum longward of this emission
line increases, as seen in the spectra of several other 
galaxies in Fig. \ref{fig8}. Note also that the spectrum of {\it g}1 is 
contaminated by the emission lines of Tol 1214--277.

In Table \ref{tab8} we show the rest-frame and observed wavelengths, fluxes 
and equivalent widths of the lines in the emission-line galaxies, and 
redshifts, derived for each line. Note that,
despite the noisy spectra, the redshifts derived from different emission lines
in the same galaxy are in very good agreement. 

All emission-line galaxies in the fields of Tol 1214--277 and Tol 65 except 
for galaxy {\it g}2, are more distant background galaxies with redshifts ranging
from $\sim$ 0.1 to $\sim$ 0.7. In particular galaxy {\it g}5,
suggested by Papaderos et al. (\cite{P99}) to be a low-surface-brightness
companion galaxy of Tol 65, is in fact a background galaxy.

The only companion galaxy to Tol 1214--277 is {\it g}2. Its spectrum is 
characterized by a flux increase 
to the blue and two weak emission lines, [O {\sc iii}] $\lambda$5007 and
H$\alpha$, with the same redshift as Tol 1214--277.
%The spectroscopic properties of the galaxy {\it g}2 suggest that
%it is in the post-starburst stage. 
%With a 
%redshift $z$ = 0.0260 and a Hubble constant $H_0$ = 75 km s$^{-1}$ Mpc$^{-1}$,
The projected linear distance between Tol 1214--277 and {\it g}2 is 14.5 kpc.

Tol 1214--277 is not unique among 
the most metal-deficient BCDs in having a companion galaxy. 
Such companions are also seen for other 
very metal-deficient BCDs such as SBS 0335--052 
(Pustilnik et al. \cite{PU97}) and HS 0822+3542 (Pustilnik et al. \cite{PU03})
at projected distances of
$\sim$ 22 kpc and $\sim$ 11 kpc, respectively. VLA 21 cm observations
of SBS 0335--052 (Pustilnik et al.
\cite{PU01}) reveal that the BCD and its companion are embedded in  
a large cloud of neutral gas implying a common origin.
It would be interesting to carry out such H {\sc i} VLA 
observations for Tol 1214--277 to check  
whether the BCD and its companion galaxy 
are also part of the same system.

\section{Conclusions \label{sum}}

Our main conclusions from the VLT spectroscopic study of the blue compact dwarf
(BCD) galaxies Tol 1214--277 and Tol 65 and the fields around these
galaxies may be summarized as follows:

1. The oxygen abundances 12 + log O/H in the brightest regions of Tol 1214--277
and Tol 65 derived from high-resolution spectra are, respectively,
7.55 $\pm$ 0.01 and 7.54 $\pm$ 0.01, or $\sim$ $Z_\odot$/24. These values
are in good agreement with previous determinations 
by Fricke et al. (\cite{F01}) and Izotov et al. (\cite{I01a}).
The high spectral resolution allows to 
separate the [N {\sc ii}] $\lambda$6583 emission line from the strong H$\alpha$
emission line and reliably determine the nitrogen abundance. We find
log N/O = --1.64 $\pm$ 0.03 and --1.60 $\pm$ 0.02 respectively, in agreement
with previous determinations by Izotov et al. 
(\cite{I01a}), and consistent with the mean value of log N/O = --1.60 obtained
by Izotov \& Thuan (\cite{IT99}) for very metal-deficient BCDs with
12 + log O/H $<$ 7.6.

2. The He mass fraction derived for several regions 
in Tol 1214--277 and Tol 65 is in the range $Y$ = 0.242 -- 0.250,
consistent with the values obtained 
by Izotov et al. (\cite{I01a}) for both galaxies, and with the primordial
He mass fraction $Y_{\rm p}$ = 0.244 -- 0.245 obtained by 
Izotov \& Thuan (\cite{IT98,IT04}) and Izotov et al. (\cite{I99}). An
exception is the He mass fraction $Y$ = 0.237 derived for regions 1 + 2
in Tol 65 because of likely 
contamination by underlying stellar He {\sc i} absorption.
However, at the 95\% confidence level,
this value is consistent with other determinations of the He abundance
in Tol 65.

3. Strong He {\sc ii} $\lambda$4686 emission, with an intensity as high 
as 5\% of that 
of the H$\beta$ emission line, and the high ionization line 
[Fe {\sc v}] $\lambda$4227 are seen in the spectrum of Tol 1214--277, 
confirming previous findings by Fricke et al.
(\cite{F01}) and Izotov et al. (\cite{I01a}). Additionally, weak 
[Ne {\sc iv}] $\lambda$4725, [Fe {\sc vi}] $\lambda$5146, $\lambda$5177,
[Fe {\sc vii}] $\lambda$5721, $\lambda$6087 emission lines are detected. 
This implies the presence of intense X-ray emission in Tol 1214--277. 
%from one or several compact sources 
%inside the brightest H {\sc ii} region of Tol 1214--277.
In particular, to produce the observed [Fe {\sc vii}] emission line 
fluxes and assuming that the X-ray sources are located in the compact region,
the X-ray luminosity of the brightest H {\sc ii} region in Tol 1214--277
should be as high as 10$^{39}$ -- 10$^{40}$ erg s$^{-1}$.

4. We find four emission-line galaxies in the field of Tol 1214--277, and
four emission-line galaxies in the field of Tol 65. Seven of these galaxies
are background star-forming galaxies with redshifts in the range 0.1 --
0.7. One emission-line galaxy in the field of Tol 1214--277 has the same
redshift as the BCD and is likely a companion galaxy at 
a projected distance of $\sim$ 14.5 kpc.

\begin{acknowledgements}
Y.I.I. acknowledges the G\"ottingen Academy of Sciences
for a Gauss professorship. He and N.G.G. have been supported by DFG grant 
436 UKR 17/22/03, by Swiss SCOPE 7UKPJ62178 grant and by grant No. 02.07/00132
of the Ukrainian fund for fundamental investigations. 
They are grateful for the hospitality of the G\"ottingen Observatory. 
%Y.I.I., N.G.G., 
P.P. and K.J.F. acknowledge support by the Volkswagen 
Foundation under grant No. I/72919. Y.I.I. and T.X.T. have been partially 
supported by NSF grant AST 02-05785. 
The research described in this publication was made possible in part by Award
No. UP1-2551-KV-03 of the U.S. Civilian Research \& Development Foundation 
for the Independent States of the Former Soviet Union (CRDF).
%Y.I.I., N.G.G. and T.X.T. acknowledge the 
%support of grant UP1-2551-KV-03 from the US Civilian Research and Development 
%Foundation (CRDF) for the independent states of the former Soviet Union.
\end{acknowledgements}


\begin{thebibliography}{}
\bibitem[1984]{A84} Aller, L. H. 1984, Physics of Thermal Gaseous Nebulae,
   Dordrecht: Reidel
\bibitem[1989]{Anders89} Anders, E., \& Grevesse, N.
  1989, Geochim.Cosmochim.Acta, 53, 197
\bibitem[1986]{C86} Campbell, A., Terlevich, R., \& Melnick, J. 1986, \mnras,
223, 811
\bibitem[1996]{DS96} Dopita, M. A., \& Sutherland, R. S. 1996, ApJS, 102, 161
%\bibitem[1996]{D96} Dufour, R. J., Esteban, C., \& Casta\~neda, H. O.
%1996, \apj, 471, L87
\bibitem[1996]{F96} Ferland, G. J. 1996, CLOUDY, Univ. of Kentucky, Dept. of
Phys. and Astron. Internal Rep.
\bibitem[1998]{F98} Ferland, G. J., Korista, K. T., Verner, D. A., 
Ferguson, J. W., Kingdon, J. B., \& Verner, E. M. 1998, PASP, 110, 761
\bibitem[2001]{F01} Fricke, K. J., Izotov, Y. I., Papaderos, P., 
Guseva, N. G., \& Thuan, T. X. 2001, \aj, 121, 169
\bibitem[1992]{G92} Garnett, D. R. 1992, \aj, 103, 1330
\bibitem[2003]{G03} Guseva, N. G., Papaderos, P., Izotov, Y. I., Green, R. F.,
Fricke, K. J., Thuan, T. X., \& Noeske, K. G. 2003, A\&A, 407, 105
\bibitem[1998]{IT98} Izotov, Y. I., \& Thuan, T. X. 1998, \apj, 500, 188
\bibitem[1999]{IT99} Izotov, Y. I., \& Thuan, T. X. 1999, \apj, 511, 639
\bibitem[2004]{IT04} Izotov, Y. I., \& Thuan, T. X. 2004, \apj, 602, 200
\bibitem[2001a]{I01a} Izotov, Y. I., Chaffee, F. H., \& Green, R. F. 2001a,
\apj, 562, 727
\bibitem[2001b]{I01b} Izotov, Y. I., Chaffee, F. H., \& Schaerer, D. 2001b,
\aap, 378, L45
\bibitem[1994]{ITL94} Izotov, Y. I., Thuan, T. X., \& Lipovetsky, V. A. 1994, 
\apj, 435, 647  
\bibitem[1997a]{ITL97} Izotov, Y. I., Thuan, T. X., \& Lipovetsky, V. A. 
1997a, \apjs, 108, 1
\bibitem[1997b]{I97b} Izotov, Y. I., Lipovetsky, V. A., Chaffee, F. H., 
Foltz, C. B., Guseva, N. G., \& Kniazev, A. Y. 1997b, \apj, 476, 698
\bibitem[1999]{I99} Izotov, Y. I., Chaffee, F. H., Foltz, C. B., et al.
1999, \apj, 527, 757
\bibitem[2004]{I04} Izotov, Y. I., Noeske, K. G., Guseva, N. G., 
Papaderos, P., Thuan, T. X., \& Fricke, K. J. 2004, \aap, 415, L27
%\bibitem[2004b]{I04b} Izotov, Y. I., Stasi\'nska, G., Guseva, N. G., \& Thuan, T. X.
%2004b, A\&A, 415, 87
%\bibitem[1995]{KF95} Kingdon, J., \& Ferland, G. J. 1995, \apj, 442, 714
\bibitem[1979]{K79} Kurucz, R. L. 1979, ApJS, 40, 1
\bibitem[1983]{K83} Kunth, D., \& Sargent, W. L. W. 1983, \apj, 273, 81
%\bibitem[1979]{L79} Lequeux, J., Rayo, J. F., Serrano, A., Peimbert, M.,
%\& Torres-Peimbert, S. 1979, \aap, 80, 155
\bibitem[1999]{L99} Lipovetsky, V. A., Chaffee, F. H., Izotov, Y. I., et al.
1999, \apj, 519, 177
\bibitem[1994]{M94} Masegosa, J., Moles, M., \& Campos-Aguilar, A. 1994,
\apj, 420, 576
\bibitem[2002]{MM02} Meynet, G., \& Maeder, A. 2002, \aap, 381, L25
\bibitem[2001]{N01} Noeske, K. G., Iglesias-P\'aramo, J., V\'ilchez, J. M.,
Papaderos, P., \& Fricke, K. J. 2001, \aap, 371, 806
\bibitem[1992]{P92} Pagel, B. E. J., Simonson, E. A., Terlevich, R. J., \&
Edmunds, M. G. 1992, \mnras, 255, 325
\bibitem[1999]{P99} Papaderos, P., Fricke, K. J., Thuan, T. X., 
Izotov, Y. I., \& Nicklas, H. 1999, \aap, 352, L57
%\bibitem[1998]{P98} Papaderos, P., Izotov, Y. I., Fricke, K. J., 
%Guseva, N. G., \& Thuan, T. X. 1998, \aap, 338, 43
\bibitem[2001]{PU01} Pustilnik, S. A., Brinks, E., Thuan, T. X., 
Lipovetsky, V. A., \& Izotov, Y. I. 2001, \aj, 121, 1413
\bibitem[1997]{PU97} Pustilnik, S. A., Lipovetsky, V. A., Izotov, Y. I.,
et al. 1997, AstL, 23, 308
\bibitem[2003]{PU03} Pustilnik, S. A., Kniazev, A. Y., Pramskij, A. G.,
 Ugryumov, A. V., \& Masegosa, J. 2003, A\&A, 409, 917
%\bibitem[1968]{R68} Robbins, R. R. 1968, \apj, 151, 511
%\bibitem[1970]{SS70} Sargent, W. L. W., \& Searle, L. 1970, \apj, 162, L155
\bibitem[2002]{S02} Schaerer, D. 2002, A\&A, 382, 28
\bibitem[2003]{S03} Schaerer, D. 2003, A\&A, 397, 527
\bibitem[1997]{S97} Schaerer, D., \& de Koter, A. 1997, A\&A, 322, 598
\bibitem[1996]{S96} Smits, D. P. 1996, \mnras, 278, 683
\bibitem[1990]{S90} Stasi\'nska G. 1990, \aaps, 83, 501
\bibitem[1997]{TI97} Thuan, T. X., \& Izotov, Y. I. 1997, \apj, 489, 623
%\bibitem[1981]{TM81} Thuan, T. X., \& Martin, G. E. 1981, \apj, 247, 823
%\bibitem[1999a]{TIF99} Thuan, T. X., Izotov, Y. I., \& Foltz, C. B. 1999a,
%\apj, 525, 105
\bibitem[1995]{TIL95} Thuan, T. X., Izotov, Y. I., \& Lipovetsky, V. A. 1995, 
\apj, 445, 108
%\bibitem[1997]{TIL97} Thuan, T. X., Izotov, Y. I., \& Lipovetsky, V. A. 1997, 
%\apj, 477, 661
%\bibitem[1999b]{TLMP99} Thuan, T. X., Lipovetsky, V. A., Martin, J.-M., 
%\& Pustilnik, S. A. 1999b, \aaps, 139, 1
\bibitem[2004]{T04} Thuan, T. X., Bauer, F. E., Papaderos, P., 
\& Izotov, Y. I., 2004, \apj, May 1
%\bibitem[1998]{V98} van Zee, L., Westphahl, D., Haynes, M., 
%\& Salzer, J. J. 1998, \aj, 115, 1000
\bibitem[1958]{W58} Whitford, A.E. 1958, \aj, 63, 201

%
\end{thebibliography}
\end{document}